\documentclass[lettersize,journal]{IEEEtran}
\usepackage{amsmath,amsfonts}
\usepackage{algorithmic}
\usepackage{array}
\usepackage[caption=false,font=normalsize,labelfont=sf,textfont=sf]{subfig}
\usepackage{textcomp}
\usepackage{stfloats}
\usepackage{url}
\usepackage{verbatim}
\usepackage{graphicx}
\def\BibTeX{{\rm B\kern-.05em{\sc i\kern-.025em b}\kern-.08em
    T\kern-.1667em\lower.7ex\hbox{E}\kern-.125emX}}
\usepackage{balance}
\usepackage{cite}

\usepackage[numbers,sort&compress]{natbib}  
\usepackage{bm}
\usepackage{algorithm}      
\usepackage{float}          
\usepackage{multirow}
\usepackage{tabularx}
\usepackage{rotating} 

\begin{document}
\title{ADSEL: Adaptive Dual Self-Expression Learning for EEG Feature Selection via Incomplete Multi-Dimensional Emotion Labels}

\author{Xueyuan~Xu, \IEEEmembership{Member, IEEE,} Tianze~Yu, Junming~Zhang, Chenrui~Liu, Wenjia~Dong, Fulin~Wei, and Li~Zhuo
\thanks{This work is funded by: National Natural Science Foundation of China under Grant No.62502019, and The Open Project of the Engineering Research Center of Autonomous Unmanned System Technology, Ministry of Education under Grant No.zzwr202511. (Corresponding author: Xueyuan Xu)}
\thanks{X.Y. Xu, T.Z. Yu, J.M. Zhang, C.R. Liu, W.J. Dong, and L. Zhuo are members of the Beijing Key Laboratory of Computational Intelligence and Intelligent System, Beijing University of Technology, Beijing, 100124, CN. (xxy@bjut.edu.cn, 23027418@emails.bjut.edu.cn, junming@emails.bjut.edu.cn, liuchenrui2023@emails.bjut.edu.cn, 23027425@emails.bjut.edu.cn, zhuoli@bjut.edu.cn)}
\thanks{X.Y. Xu is also a member of the Engineering Research Center of Autonomous Unmanned System Technology, Ministry of Education, Hefei, 230601, CN.}
\thanks{F.L. Wei is a member of the School of Artificial Intelligence, Anhui University, Hefei, 230601, CN. (weifulin@ahu.edu.cn)}}

\maketitle

\begin{center}
\small
This work has been submitted to the IEEE for possible publication. 
Copyright may be transferred without notice, after which this version may no longer be accessible.
\end{center}

\begin{abstract}
Electroencephalography (EEG) based multi-dimension emotion recognition has attracted substantial research interest in the field of human-computer interaction. However, the high dimensionality of EEG features, coupled with limited sample sizes, frequently leads to classifier overfitting and high computational complexity. Feature selection constitutes a critical strategy for mitigating these challenges. However, most existing EEG feature selection methods assume complete multi-dimensional emotion labels. In practice, open acquisition environment and the inherent subjectivity of emotion perception often result in incomplete label data, which can compromise model generalization. Additionally, existing feature selection methods for handling incomplete multi-dimensional labels primarily focus on correlations among various dimensions during label recovery, neglecting the correlation between samples in the label space and their interaction with various dimensions. To address these issues, we propose a novel incomplete multi-dimensional emotion feature selection framework integrating Adaptive Dual Self-Expression Learning (ADSEL) with least squares regression. ADSEL could establish a bidirectional pathway between sample-level and dimension-level self-expression learning processes within the label space. It could facilitate the cross-sharing of learned information between these processes, enabling the simultaneous exploitation of effective information across both samples and dimensions for label reconstruction. Consequently, ADSEL could enhance label recovery accuracy and effectively identifies the optimal EEG feature subset for multi-dimensional emotion recognition. ADSEL was evaluated against fourteen state-of-the-art feature selection methods on three public EEG datasets with multi-dimensional emotion labels. Experimental results demonstrate that ADSEL could achieve superior performance under conditions of partial label absence.
\end{abstract}


\section{Introduction}

\IEEEPARstart{I}{n} recent years, as a key component of affective interaction, research on emotion recognition has received extensive attention in the field of human-computer interaction. \cite{11494362,Mitchell17092025}. Fundamental affective representation models for emotion recognition fall into two main categories: discrete and dimensional models \cite{chen2025emotion}. Discrete models define a limited set of basic emotions, with complex emotions formed by combining them \cite{chen2025emotion}. While intuitive, they often lack the granularity needed for computational analysis \cite{Alarcao2019ERES}. In contrast, dimensional models map affective states onto multi-dimensional spaces, offering greater flexibility, expressiveness, and support for modeling dynamic emotional processes \cite{Alarcao2019ERES}. Among dimensional models, the Valence-Arousal (VA) model and the Valence-Arousal-Dominance (VAD) \cite{mehrabian1996pleasure} model are widely adopted.

Electroencephalography (EEG) signals have been proven particularly valuable in dimensional emotion recognition research owing to their high temporal resolution, non-invasiveness, and relative simplicity of acquisition \cite{11369788,10620218}, and it has attracted substantial interest in the field of brain-machine interface. To effectively characterize the non-stationary and nonlinear nature of EEG data associated with distinct affective states, researchers have employed diverse feature extraction methodologies. These include, but are not limited to, Non-Stationary Index (NSI) \cite{kroupi2011eeg}, Shannon Entropy (SHE) \cite{lin1991divergence}, and Higher-Order Crossing (HOC) \cite{Petrantonakis2010HOC}.


Advancements in EEG acquisition, particularly the adoption of high-density electrode arrays, have significantly increased EEG feature dimensionality \cite{Becker2020HRE}. However, EEG applications often face limited sample sizes, increasing susceptibility to the curse of dimensionality. In the emotion recognition task, this elevates risks of model overfitting/underfitting and substantial computational costs \cite{wang2020emotion}. Feature selection is pivotal to address these challenges: it identifies the most salient information related to emotional states within high-dimensional spaces, eliminating redundancy and noise while preserving discriminative neural signatures. This process enhances emotion recognition model generalizability and interpretability \cite{jenke2014taffc}.

EEG feature selection methods are broadly categorized into three paradigms: filter, wrapper, and embedded methods \cite{hou2011JELSR}. Filter methods assess feature relevance using statistical or information-theoretic criteria, but often yield suboptimal subsets irrespective of the classifier's performance \cite{zhang2019review}. To address this issue, wrapper methods evaluate subsets based directly on classifier accuracy, typically achieving superior results \cite{hou2011JELSR}. However, their exhaustive search can be computationally intensive \cite{zhang2019review}. As an alternative, embedded methods integrate selection within the model training process itself. Previous research \cite{xu2020fsorer, GRMOR2021taffc}
demonstrate their efficacy for emotion recognition.

Most feature selection methods for EEG-based emotion recognition universally presume the availability of fully annotated and reliable training labels. However, in practical deployment scenarios, label absence constitutes a prevalent and often unavoidable challenge, stemming from open acquisition environments and inherent inconsistencies in subjective emotional annotation. This missing-label condition significantly degrades the performance of existing EEG feature selection techniques, consequently impairing emotion recognition accuracy. 



To address the challenge of missing labels in EEG-based emotion recognition, this paper proposes a novel feature selection framework termed Adaptive Dual Self-Expression Learning (ADSEL). Unlike conventional methods that rely on complete supervision or treat label recovery and feature selection as separate processes, ADSEL integrates them within a unified least squares regression architecture. Its core innovation lies in establishing a bidirectional pathway between sample-level and dimension-level self-expression learning within the label space. Specifically, the model jointly learns two coefficient matrices. One captures dependencies among samples, while the other captures correlations among emotion dimensions. Cross-information sharing between the two matrices is then enforced to facilitate mutual enhancement. This dual mechanism enables the framework to simultaneously exploit structural regularities across instances and semantic correlations across labels, thereby achieving robust label reconstruction even under high missing rates. Moreover, by coupling this adaptive label learning module with global feature redundancy minimization and graph-based manifold regularization, ADSEL ensures that the final selected EEG feature subset is not only discriminative but also compact and locally structure-preserving. Consequently, the method effectively mitigates the adverse effects of incomplete labels and identifies stable features for multidimensional affective states.

Furthermore, the contributions of this article are as follows:
\begin{enumerate}
  \item[$\bullet$]A novel EEG feature selection method is proposed to address the challenge of incomplete labels in emotion recognition by seamlessly integrating it within an extended ADSEL model framework. This framework collaboratively integrating information from both the samples and dimensions in the label space to reconstruct absent emotional label, explicitly mitigates the adverse impacts of missing label on feature selection.
  
  \item[$\bullet$] To tackle the optimization difficulties inherent in ADSEL, an efficient alternative optimization approach is devised. This strategy guarantees algorithmic convergence and ensures convergence to a local optimum.
  
  \item[$\bullet$] The efficacy of ADSEL for incomplete multi-dimensional emotion feature selection is rigorously evaluated on three established multi-dimensional affective datasets with EEG recordings (DEAP, DREAMER, and AMIGOS). Comparative analysis against fourteen state-of-the-art feature selection methods demonstrates that EEG feature subsets selected by ADSEL achieve statistically superior emotion recognition performance across all six evaluation metrics.
\end{enumerate}

\section{Related Work}
\subsection{Notations}
Consistent notation is employed throughout this work. Vectors appear in lowercase boldface ($\mathbf{x}$, $\mathbf{y}$, ...), and matrices in uppercase ($\mathbf{X}$, $\mathbf{Y}$, ...). Matrix transpose and trace are denoted by $\mathbf{X}^{T}$ and $\operatorname{Tr}\mathbf{(X)}$, respectively. The Hadamard product is symbolized by $\odot$. The Frobenius norm and $l_{2,1}$ norm of a matrix $\mathbf{X}$ are defined as:
%
%
The Frobenius norm and $l_{2,1}$ norm of a matrix $\mathbf{X}$ are defined as $\|\mathbf{X}\|_{F}=\sqrt{\sum_{i=1}^{d} \sum_{j=1}^{n} x_{i j}^{2}}=\sqrt{Tr\left(\mathbf{X}^{T} \mathbf{X}\right)}$ and $\|\mathbf{X}\|_{2,1}=\sum_{i=1}^{d} \sqrt{\sum_{j=1}^{n} x_{i, j}^2}=\sum_{i=1}^{d}\left\|\mathbf{x}_{i,:}\right\|_2$.

The EEG feature matrix $\mathbf{X} \in \mathbb{R}^{d \times n}$ contains $d$ features, where row vector $\mathbf{x}_i \in \mathbb{R}^{1 \times n}$ ($i = 1, ..., d$) spans $n$ samples. Emotional states are encoded in the label matrix $\mathbf{Y} \in{\{0,1\}}^{n\times k}$, representing $k$ emotion dimensions across $n$ instances. The all-one vector $\mathbf{1}_n =(1,1 \ldots, 1)^{T} \in \mathbb{R}^{n \times 1}$ 
and identity matrix $\mathbf{I_n}\in \mathbb{R}^{n \times n}$ are standard.

\subsection{EEG feature selection methods}

Owing to diverse search strategies, feature selection methodologies are commonly classified into three primary categories: filter methods, wrapper methods, and embedded methods \cite{zhang2019review}.

Filter methods evaluate features based on intrinsic statistical properties, such as correlation coefficients or information-theoretic measures, ranking features independently of the learning model \cite{saeys2007review}. However, they may overlook features whose discriminative power emerges only in combination. Wrapper methods adopt classifier performance to evaluate feature subsets, employing distinct search strategies like evolutionary computation \cite{nakisa2018evolutionary} to explore combinations. Their main drawback is high computational cost due to iterative model training \cite{saeys2007review}.

Embedded methods integrate feature selection directly into model training, simultaneously optimizing model parameters and feature importance \cite{saeys2007review}. Least Squares Regression (LSR) is a widely adopted foundation \cite{wu2019FSOR,GRMOR2021taffc,10403843}. LSR-based algorithms learn a projection matrix $\mathbf{W}$, quantifying feature significance by $\{\|\mathbf{{w}}_{1}\|_{2},..., \|\mathbf{{w}}_{d}\|_{2}\}$ \cite{yang2019unsupervised} to select discriminative subsets. To improve the robustness and compactness of the selected features, several extensions have been proposed. Representative examples include Feature Selection with Orthogonal Regression (FSOR)~\cite{xu2020fsorer}, which introduces orthogonal regression to enhance feature discriminability; Global Redundancy Minimization Orthogonal Regression (GRMOR)~\cite{GRMOR2021taffc}, which explicitly models global feature redundancy via correlation constraints; Feature Selection method via Manifold Regularization (MDFS)~\cite{zhang2019manifold}, which performs multi-label feature selection via dynamic graph regularization; and Global Relevance and Redundancy Optimization (GRRO)~\cite{zhang2020multi}, which leverages global redundancy and relevance optimization. Other extensions incorporate sparse regularization \cite{lee2017scls} or leverage local and global label relevance \cite{10403843} to further improve feature selection performance.


\subsection{Incomplete multi-label feature selection}

 With missing labels common in practice, feature selection for incomplete labels is gaining attention. For example, Zhu et al. \cite{zhu2018multi} proposed Multi-Label Feature Selection with Missing Labels method (MLMLFS), recovering labels via linear regression and using $l_{2,p}$ regularization for feature selection, but it ignores label correlation and relies on linearity. Yin et al. \cite{yin2024feature} proposed Feature Selection for Multilabel classification with Missing labels (FSMML) employs a multi-scale fuzzy roughness framework, mapping data to multi-scale space. It balances feature correlation/redundancy via label distribution reconstruction and fuzzy uncertainty fusion to correct label bias, but incurs high complexity and weak interpretability. Meanwhile, Xu et al. \cite{xu2024wsel} proposed Weighted Self-expression Learning (WSEL), which presented a feature-centric method tailored for high-dimensional data that captures latent structures without explicit label imputation, though it remains sensitive to noise and outliers due to its reliance on the quality of the feature space. Dai et al. \cite{dai2025multi} proposed Weak-label Fusion Fuzzy Discernibility Pair (WFDP) circumvents explicit recovery, using fuzzy rough sets to model feature discrimination via sample-pair distinguishability under missing labels.
 
\begin{figure*}[!t]
\centering
\includegraphics[width=0.8\textwidth]{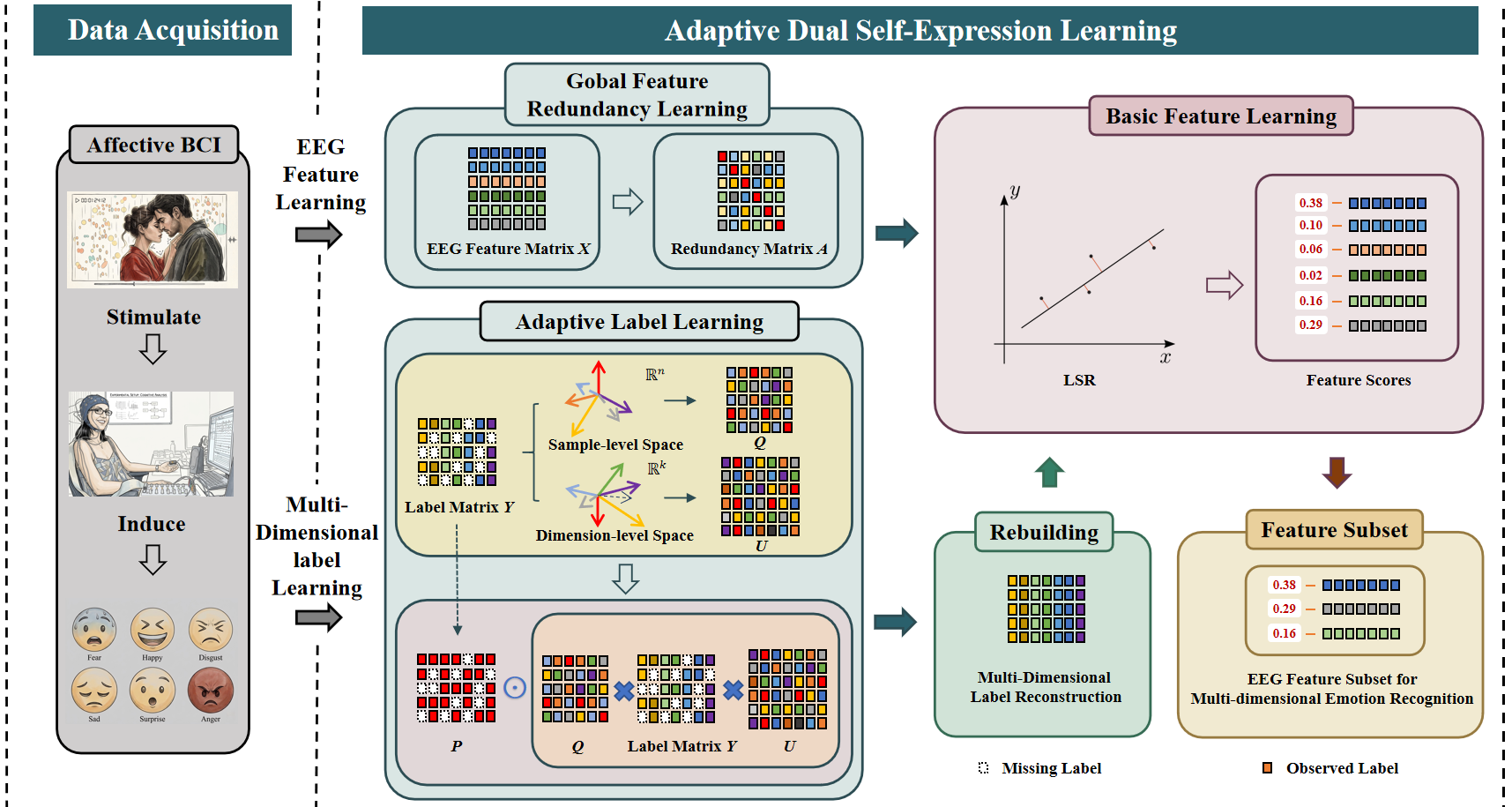}
\caption{The proposed ADSEL framework concludes the following three sections: (a) Basic feature learning; (b) Adaptive label learning; (c) Global feature redundancy learning. The overall workflow is as follows: First, incomplete multi-dimensional emotion data undergoes label reconstruction via adaptive dual self-representation learning. Second, redundancy among EEG features is evaluated through global redundancy learning. Finally, by integrating the reconstructed label data, global redundancy information, and least squares regression, a compact and robust EEG emotion feature subset is constructed.}
\label{Framework_efsmder}
\end{figure*}

\section{Problem formulation}
This section presents ADSEL, a novel EEG emotion feature selection approach. ADSEL is designed to obtain non-redundant and relevant feature subsets for incomplete multi-dimensional affective computing. The ADSEL framework is defined as follows:
\begin{equation}
\min _{\mathbf{W}, \mathbf{Q}, \mathbf{U}} F(\mathbf{X}, \mathbf{W}, \mathbf{Q}, \mathbf{Y}, \mathbf{U})+\lambda C(\mathbf{Q}, \mathbf{Y}, \mathbf{U})+\gamma \Omega(\mathbf{W})
\end{equation}
where the projection matrix and two coefficient matrices are represented by the variables $\mathbf{W}$, $\mathbf{Q}$, and $\mathbf{U}$, respectively. $\gamma$ and $\lambda$ are both tradeoff parameters. The basic feature learning function, adaptive label learning function, and global feature redundancy learning function are denoted by the symbols $F$, $C$, and $\Omega$, respectively. The next subsections will introduce the definitions of $F$, $C$, and $\Omega$.

\subsection{Basic feature learning}
LSR is adopted to model the relationship between the EEG feature matrix $\mathbf{X}$ and the dual self-expression matrix $\mathbf{Q} \mathbf{Y} \mathbf{U}$. For effective EEG feature selection, the projection matrix $\mathbf{W}$ is regularized using the $\ell_{2,1}$-norm to induce structured sparsity and select discriminative features. Consequently, the functional term is formulated as:

\begin{equation}
\begin{aligned}
\label{equ_framework_lsr} F(\mathbf{X}, \mathbf{W}, \mathbf{Q}, \mathbf{Y}, \mathbf{U}) = & \left\|\mathbf{X}^T \mathbf{W}+\mathbf{1}_{n} \mathbf{b}^T-\mathbf{Q} \mathbf{Y} \mathbf{U}\right\|_F^2\\&+\delta\|\mathbf{W}\|_{2,1} \\
\end{aligned}
\end{equation}
where $\mathbf{b} \in \mathbb{R}^{k \times 1}$ denotes the bias vector, $\mathbf{W} \in \mathbb{R}^{d \times k}$ represents the projection matrix, $\mathbf{Q} \in \mathbb{R}^{n \times n}$ and $\mathbf{U} \in \mathbb{R}^{k \times k}$ are the coefficient matrices for the sample and dimension perspectives in the label space, respectively, and $\delta$ ($\delta > 0$) serves as a trade-off parameter.

\subsection{Adaptive label learning}

The dual correlation inherent within the label space is leveraged: both samples and dimensions exhibit linear dependencies: individual samples can be represented by others, and each dimension is jointly expressible by all dimensions. Inspired by this dual correlation, a dual self-expression model is proposed for label matrix reconstruction. This model synergistically exploits the structural information within samples and semantic information between dimensions in the label space through its dual linear formulation, thereby achieving high-precision and robust label recovery. The label reconstruction formula using dual self-expression is as follows:

\begin{equation}
\label{equ_framework_se}
\min _{\mathbf{Q}, \mathbf{U}} \| (\mathbf{Y}- \mathbf{Q} \mathbf{Y} \mathbf{U})\|_{F}^{2} 
\end{equation}

To eliminate the impact of missing labels and guide recovery using known labels, an missing-indicator matrix $\mathbf{P} \in \mathbb{R}^{n \times k}$ is introduced, defined as follows:

\begin{equation}
\mathbf{P}_{i,j}= \begin{cases}1 & \text { if } i \text {-th instance exists in } j \text {-th dimensional label; } \\ 0 & \text { otherwise. }\end{cases}
\end{equation}

Then, $\mathbf{P}$ is introduced into the dual self-expression model as follows:

\begin{equation}
\begin{aligned}
\label{equ_framework_se_1}
\min _{\mathbf{Q}, \mathbf{U}} \|\mathbf{P} \odot (\mathbf{Y}- \mathbf{Q} \mathbf{Y} \mathbf{U})\|_{F}^{2} +\alpha\|\mathbf{U}\|_{2,1}\\
\end{aligned}
\end{equation}

To ensure that each label in the label space is reconstructed primarily by its most relevant labels, an $l_{2,1}$-norm constraint is enforced on the self-expression coefficient matrix $\mathbf{U}$.

Furthermore, to preserve the consistency of local geometric structures between the dual self-expression space $\mathbf{Q} \mathbf{Y} \mathbf{U}$ and the sample space $\mathbf{X}$, a graph-based manifold regularizer is incorporated. The graph Laplacian matrix $\mathbf{L}_X\in {\mathbb{R}^{n\times n}}$ is formally defined as $\mathbf{L}_{X} = \mathbf{G} - \mathbf{S}$. Here, $\mathbf{S}$ denotes the affinity graph constructed for $\mathbf{Y}$, and $\mathbf{G}$ is the corresponding degree matrix with diagonal elements computed as $\mathbf{G}_{i,i} = \sum_{j=1}^{n} \mathbf{S}_{i,j}$. The affinity graph $\mathbf{S}$ is generated using a standard heat kernel, where the similarity $\mathbf{S}_{i,j}$ between two label vectors $\mathbf{y}_{i}$ and $\mathbf{y}_{j}$ is given by:

\begin{equation}
\mathbf{S}_{i,j}=\left\{\begin{array}{ll}
\exp \left(-\frac{\left\|\mathbf{y}_{i}-\mathbf{y}_{j}\right\|^{2}}{\sigma^{2}}\right) & 
  \substack{
    \mathbf{y}_{i} \in \mathcal{N}_{q}\left(\mathbf{y}_{j}\right)  \\ 
    \text{or} \\ 
    \mathbf{y}_{j} \in \mathcal{N}_{q}\left(\mathbf{y}_{i}\right)
  } \\
0 & \text{otherwise}
\end{array}\right.
\end{equation}

Here, $\sigma$ signifies the graph construction parameter, and $\mathcal{N}_{p}\left(\mathbf{y}_{j}\right)$ corresponds to the $q$-nearest-neighbor set of the label vector $\mathbf{y}_{j}$ in the label space.

Thus, by incorporating the graph-based manifold regularizer into the adaptive dual self-expression framework, the following formulation for term $C$ is obtained :

\begin{equation}
\begin{aligned}
\label{equ_framework_R}
C(\mathbf{Q}, \mathbf{Y}, \mathbf{U})= &\|\mathbf{P} \odot(\mathbf{Y}- \mathbf{Q} \mathbf{Y} \mathbf{U})\|_{F}^{2}  +\alpha\|\mathbf{U}\|_{2,1}\\&+\beta Tr\left((\mathbf{Q} \mathbf{Y} \mathbf{U})^T \mathbf{L}_X (\mathbf{Q} \mathbf{Y} \mathbf{U})\right)\\
&\text { s.t. } \mathbf{Q} \geq 0,\mathbf{U} \geq 0
\end{aligned}
\end{equation}
where $\beta$ and $\alpha$ represent trade-off parameters.

\subsection{Global feature redundancy learning}

Building on prior work \cite{GRMOR2021taffc}, a global feature redundancy matrix $\mathbf{A}$ is proposed to quantify paired redundancy among EEG features. Each element $\mathbf{A}_{i, j}$ quantifies the squared cosine similarity between centralized feature vectors:

\begin{equation}
\label{equ_red_matrix}
\mathbf{A}_{i, j}=\left(\mathbf{O}_{i, j}\right)^{2}=\left(\frac{\mathbf{f}_{i}^{T} \mathbf{f}_{j}}{\left\|\mathbf{f}_{i}\right\|\left\|\mathbf{f}_{j}\right\|}\right)^{2}
\end{equation}
where $\mathbf{f}_{i}\in {\mathbb{R}^{n\times 1}}$ and $\mathbf{f}_{j}\in {\mathbb{R}^{n\times 1}}$ correspond to the centered representations of EEG features $\mathbf{x}_{i}$ and $\mathbf{x}_{j}$ ($i,j=1,2,...,\textit{d}$). Accordingly, the global feature redundancy minimization term is defined by:

\begin{equation}
\begin{aligned}
\label{equ_framework_theta}
\Omega(\mathbf{W}) = \operatorname{Tr}\left(\mathbf{W}^T \mathbf{A} \mathbf{W}\right)
\end{aligned}
\end{equation}

\subsection{The final objective function of ADSEL}

Combining Eqs.~\eqref{equ_framework_lsr}, \eqref{equ_framework_R}, and \eqref{equ_framework_theta}, the proposed ADSEL is formulated as:

\begin{equation}
\begin{aligned}
\label{equ_framework}
\min _{\mathbf{b}, \mathbf{W}, \mathbf{Q}, \mathbf{U}}&\left\|\mathbf{X}^T \mathbf{W}+\mathbf{1}_{n} \mathbf{b}^T-\mathbf{Q} \mathbf{Y} \mathbf{U}\right\|_F^2+\delta\|\mathbf{W}\|_{2,1}\\&+\lambda\|\mathbf{P} \odot(\mathbf{Y}-\mathbf{Q} \mathbf{Y} \mathbf{U})\|_F^2+\alpha\|\mathbf{U}\|_{2,1} \\&+\beta \operatorname{Tr}\left((\mathbf{Q} \mathbf{Y} \mathbf{U})^T \mathbf{L}_X (\mathbf{Q} \mathbf{Y} \mathbf{U})\right)+\mu \operatorname{Tr}\left(\mathbf{W}^T \mathbf{A} \mathbf{W}\right) \\
&\text { s.t. } \mathbf{Q} \geq 0,\mathbf{U} \geq 0
\end{aligned}
\end{equation}
where $\lambda$, $\beta$, $\alpha$, $\mu$, and $\delta$ are regularization parameters. The framework of ADSEL is shown in Fig.~\ref{Framework_efsmder}.

\section{Optimization Strategy} \label{Optimization Strategy}

Setting the partial derivative of Eq.~\eqref{equ_framework} with respect to $\mathbf{b}$ to zero allows us to solve for the bias term, yielding $\mathbf{b} = \frac{1}{n} \left( \mathbf{U}^{T} \mathbf{Y}^{T} \mathbf{Q}^{T} \mathbf{1}_{n} - \mathbf{W}^{T} \mathbf{X} \mathbf{1}_{n} \right)$. Substituting this expression for $\mathbf{b}$ back into Eq.~\eqref{equ_framework} transforms the original optimization problem into the following form:

\begin{equation}
\begin{aligned}
\label{equ_framework_re}
\min _{\mathbf{W},\mathbf{Q},\mathbf{U}}&\left\|\mathbf{H} \mathbf{X}^T \mathbf{W} -\mathbf{H} \mathbf{Q} \mathbf{Y} \mathbf{U}\right\|_F^2+\delta\|\mathbf{W}\|_{2,1}\\&+\lambda\|\mathbf{P} \odot(\mathbf{Y}-\mathbf{Q} \mathbf{Y} \mathbf{U})\|_F^2+\alpha\|\mathbf{U}\|_{2,1}\\& +\beta \operatorname{Tr}\left((\mathbf{Q} \mathbf{Y} \mathbf{U})^T \mathbf{L}_{\mathbf{X}}(\mathbf{Q} \mathbf{Y} \mathbf{U})\right)+\mu \operatorname{Tr}\left(\mathbf{W}^T \mathbf{A} \mathbf{W}\right) \\
&\text { s.t. } \mathbf{Q} \geq 0,\mathbf{U} \geq 0
\end{aligned}
\end{equation}

We employ an alternating iterative update technique to solve for the three variables ($\mathbf{W}$, $\mathbf{Q}$, $\mathbf{U}$) in Eq.~\eqref{equ_framework_re}. The specific procedure is detailed below:

\subsection{Update $\mathbf{W}$ by fixing $\mathbf{Q}$ and $\mathbf{U}$}
When $\mathbf{Q}$ and $\mathbf{U}$ is fixed and we remove the irrelevant terms, we obtain the following function about $\mathbf{W}$:

\begin{equation}
\begin{aligned}
\label{equ_w_l}
\mathcal{L}\left(\mathbf{W}\right) = &\left\|\mathbf{H} \mathbf{X}^T \mathbf{W} -\mathbf{H} \mathbf{Q} \mathbf{Y} \mathbf{U}\right\|_F^2+\mu \operatorname{Tr}\left(\mathbf{W}^T \mathbf{A} \mathbf{W}\right)\\&+\delta\|\mathbf{W}\|_{2,1}
\end{aligned}
\end{equation}

By taking the partial derivative of $\mathcal{L}\left(\mathbf{W}\right)$ w.r.t.$\mathbf{W}$, we could get
\begin{equation}
\label{equ_w_pd}
\frac{\partial \mathcal{L}\left(\mathbf{W}\right)}{\partial \mathbf{W}} = 2\mathbf{X}\mathbf{H}\mathbf{X}^T\mathbf{W}-2\mathbf{X}\mathbf{H}\mathbf{Q}\mathbf{Y}\mathbf{U}+2\mu\mathbf{A}\mathbf{W}+2\delta\mathbf{D}\mathbf{W}
\end{equation}
where $\mathbf{D}$ is a diagonal matrix and the element of $\mathbf{D}$ is calculated by $\mathbf{D}_{i,i}=\frac{1}{2 \sqrt{\mathbf{W}_i^T \mathbf{W}_i+\mathbf{\epsilon}}}(\mathbf{\epsilon} \rightarrow 0)$.

Hence, the optimal solution $\mathbf{W}$ can be updated as follows:
\begin{equation}
\label{equ_w_sol}
\mathbf{W} = (\mathbf{X}\mathbf{H}\mathbf{X}^T+\mu\mathbf{A}+\delta\mathbf{D})^{-1}(\mathbf{X}\mathbf{H}\mathbf{Q}\mathbf{Y}\mathbf{U})
\end{equation}

\subsection{Update $\mathbf{Q}$ by fixing $\mathbf{W}$ and $\mathbf{U}$}

When $\mathbf{W}$ and $\mathbf{U}$ are fixed, we derive the corresponding Lagrange function by eliminating irrelevant terms and introducing a Lagrange multiplier $\mathbf{\Phi}$ for $\mathbf{Q} \geq 0$:

\begin{equation}
\begin{aligned}
\label{equ_q_pd}
\mathcal{L}\left(\mathbf{Q}\right)=&\left\|\mathbf{H} \mathbf{X}^T \mathbf{W} -\mathbf{H} \mathbf{Q} \mathbf{Y} \mathbf{U}\right\|_F^2+\lambda\|\mathbf{P} \odot(\mathbf{Y}-\mathbf{Q}\mathbf{Y}\mathbf{U})\|_F^2 \\
&+\beta \operatorname{Tr}\left((\mathbf{Q} \mathbf{Y} \mathbf{U})^T \mathbf{L}_{\mathbf{X}}(\mathbf{Q} \mathbf{Y} \mathbf{U})\right) +\operatorname{Tr}\left(\mathbf{\Phi}^T \mathbf{Q}\right) \\
\end{aligned}
\end{equation}

Taking the derivative of $\mathcal{L}(\mathbf{Q})$ with respect to $\mathbf{Q}$ yields

\begin{equation}
\begin{aligned}
\label{equ_q_derivation}
\frac{\partial \mathcal{L}}{\partial \mathbf{Q}} =& 
2\mathbf{H}\mathbf{Q}\mathbf{Y}\mathbf{U}\mathbf{U}^T\mathbf{Y}^T - 2\mathbf{H}\mathbf{X}^T\mathbf{W}\mathbf{U}^T\mathbf{Y}^T \\
&+ 2\lambda \left[\mathbf{P} \odot (\mathbf{Q}\mathbf{Y}\mathbf{U} - \mathbf{Y})\right]\mathbf{U}^T\mathbf{Y}^T \\
&+ 2\beta \mathbf{L}_{\mathbf{X}} \mathbf{Q}\mathbf{Y}\mathbf{U}\mathbf{U}^T\mathbf{Y}^T + \mathbf{\Phi}
\end{aligned}
\end{equation}

According to the Karush-Kuhn-Tucker (KKT) complementary condition $\mathbf{\Phi}_{i,j}\mathbf{Q}_{i,j} = 0$, $\mathbf{W}$ can be updated as follows:

\begin{equation}
\label{equ_q_sol}
\mathbf{Q} \leftarrow \mathbf{Q} \odot \frac{\mathbf{H} \mathbf{X}^T \mathbf{W} \mathbf{R}+\lambda (\mathbf{P} \odot \mathbf{Y}) \mathbf{R}}{\mathbf{H} \mathbf{M} \mathbf{R}+\beta \mathbf{L}_{\mathbf{X}} \mathbf{M} \mathbf{R}+\lambda \mathbf{P}\odot(\mathbf{M}) \mathbf{R}}
\end{equation}
where $\mathbf{M}=\mathbf{Q}\mathbf{Y}\mathbf{U}$ and $\mathbf{R} = \mathbf{U}^T \mathbf{Y}^T$.


\subsection{Update $\mathbf{U}$ by fixing $\mathbf{W}$ and $\mathbf{Q}$}
When $\mathbf{W}$ and $\mathbf{Q}$ are fixed, we have the following Lagrange function by removing irrelevant terms and introducing a Lagrange multiplier $\mathbf{\Psi}$ for $\mathbf{U} \geq 0$:

\begin{equation}
\begin{aligned}
\label{equ_u_pd}
\mathcal{L}\left(\mathbf{U}\right)=&\left\|\mathbf{H} \mathbf{X}^T \mathbf{W} -\mathbf{H} \mathbf{Q} \mathbf{Y} \mathbf{U}\right\|_F^2+\lambda\|\mathbf{P} \odot(\mathbf{Y}-\mathbf{Q}\mathbf{Y}\mathbf{U})\|_F^2 \\
&+\beta \operatorname{Tr}\left((\mathbf{Q} \mathbf{Y} \mathbf{U})^T \mathbf{L}_{\mathbf{X}}(\mathbf{Q} \mathbf{Y} \mathbf{U})\right) +\operatorname{Tr}\left(\mathbf{\Psi}^T \mathbf{U}\right) \\& +\alpha\|\mathbf{U}\|_{2,1}
\end{aligned}
\end{equation}

Taking the derivative of $\mathcal{L}(\mathbf{U})$ with respect to $\mathbf{U}$ yields

\begin{equation}
\begin{aligned}
\label{equ_u_derivation}
\frac{\partial \mathcal{L}}{\partial \mathbf{U}} =& 
2\mathbf{Y}^T\mathbf{Q}^T\mathbf{H}\mathbf{Q}\mathbf{Y}\mathbf{U} - 2\mathbf{Y}^T\mathbf{Q}^T\mathbf{H}\mathbf{X}^T\mathbf{W} \\
&+ 2\lambda \mathbf{Y}^T\mathbf{Q}^T\left[\mathbf{P} \odot (\mathbf{Q}\mathbf{Y}\mathbf{U} - \mathbf{Y})\right] \\
&+ 2\beta \mathbf{Y}^T\mathbf{Q}^T \mathbf{L}_{\mathbf{X}} \mathbf{Q}\mathbf{Y}\mathbf{U}
+ 2\alpha \mathbf{V}\mathbf{U} + \mathbf{\Psi}
\end{aligned}
\end{equation}
where $\mathbf{V}$ is a diagonal matrix and the element of $\mathbf{V}$ is calculated by $\mathbf{V}_{i,i}=\frac{1}{2 \sqrt{\mathbf{U}_i^T \mathbf{U}_i+\mathbf{\epsilon}}}(\mathbf{\epsilon} \rightarrow 0)$.

Via KKT condition $\mathbf{\Psi}_{i,j}\mathbf{U}_{i,j} = 0$, $\mathbf{U}$ can be updated as:

\begin{equation}
\label{equ_u_sol}
\mathbf{U} \leftarrow \mathbf{U} \odot \frac{\mathbf{N} \mathbf{H} \mathbf{X}^T \mathbf{W}+\lambda \mathbf{N} (\mathbf{P} \odot \mathbf{Y})}{\mathbf{N} \mathbf{H} \mathbf{M}+\beta \mathbf{N} \mathbf{L}_{\mathbf{X}} \mathbf{M} +\lambda \mathbf{N} \mathbf{P}\odot(\mathbf{M})+\alpha \mathbf{V} \mathbf{U}}
\end{equation}
where $\mathbf{M}=\mathbf{Q}\mathbf{Y}\mathbf{U}$ and $\mathbf{N}=\mathbf{Q}^T\mathbf{Y}^T$.

\begin{algorithm}[H]
\caption{ADSEL }
\label{ADSEL}
\begin{algorithmic}[1]
\STATE \textbf{Input}: EEG feature matrix $\mathbf{X}\in {\mathbb{R}^{d\times n}}$, incomplete multi-dimensional emotional label matrix $\mathbf{Y}\in {\mathbb{R}^{n\times k}}$, and missing-indicator matrix $\mathbf{P}\in {\mathbb{R}^{n\times k}}$.
\STATE \textbf{Parameter}: $\alpha$, $\beta$, $\lambda$, $\mu$, $\delta$.
\STATE \textbf{Output}: Ranked EEG features.
\STATE Initial $\mathbf{H} ={{\mathbf{I}}_{n}}-\frac{1}{n}{{\mathbf{1}}_{n}}{{\mathbf{1}}_{n}}^{T}$. Initial $\mathbf{W}$, $\mathbf{Q}$, and $\mathbf{U}$ randomly.
\REPEAT
\STATE Update $\mathbf{D}$ via $\mathbf{D}_{i,i}=\frac{1}{2 \sqrt{\mathbf{W}_i^T \mathbf{W}_i+\epsilon}}$;
\STATE Update $\mathbf{V}$ via $\mathbf{V}_{i,i}=\frac{1}{2 \sqrt{\mathbf{U}_i^T \mathbf{U}_i+\epsilon}}$;
\STATE Update $\mathbf{W}$  via Eq.~\eqref{equ_w_sol};
\STATE Update $\mathbf{Q}$  via Eq.~\eqref{equ_q_sol};
\STATE Update $\mathbf{U}$  via Eq.~\eqref{equ_u_sol};
\UNTIL{Convergence;}
\RETURN $\mathbf{W}$ for EEG feature selection;
\STATE Sort the EEG features by $ \|\mathbf{w}_{i}\|_{2}$.
\end{algorithmic}
\end{algorithm}

Algorithm~\ref{ADSEL} delineates the specific optimization procedures for Eq.~\eqref{equ_framework}. Within the context of multi-dimensional emotion recognition with incomplete labels, the discriminative abilities of each EEG feature is evaluated based on the weighting matrix $\mathbf{W}$. This process ultimately yields an optimal EEG feature subset characterized by both high informativeness and minimal redundancy.

\section{Experimental Details}
This section delineates the experimental framework, including data sets, EEG feature extraction and experimental setup.

\subsection{Dataset description}

To validate the ADSEL model, this study leverages three publicly available EEG datasets collected under standardized emotion-elicitation paradigms: DREAMER \cite{DREAMER2018jbhi}, DEAP \cite{koelstra2011deap}, and AMIGOS \cite{miranda2018amigos}. All datasets are annotated according to the VAD model. The detailed information is shown in Table~\ref{tab:comp_datasets}.

%
%
%


\subsection{EEG feature extraction}
Raw EEG signals underwent standardized preprocessing including 1–50 Hz band-pass filtering to remove noise and baseline drift, followed by Independent Component Analysis for artifact removal. Subsequently, feature extraction was performed on entire trials as individual samples, without any segmentation to artificially expand the sample size, thus preserving the independence of each sample for subsequent analyses. Building upon prior research in EEG-based emotion recognition \cite{jenke2014taffc,xu2020fsorer}, thirteen distinct categories of EEG features were extracted for multi-dimensional affective computing:  \textbf{(1) Time domain:} NSI \cite{kroupi2011eeg}, HOC \cite{Petrantonakis2010HOC}, Spectral Entropy (SPE), SHE \cite{lin1991divergence}, C0 complexity (C0) \cite{zhou2008study}, and Absolute Power ($AP$); \textbf{(2) Frequency domain:} the Absolute Power ratio of the theta band to the beta band ($AP_{\beta}/AP_{\theta}$), and Differential Entropy (DE); \textbf{(3) Joint time-frequency domain:} Amplitude of the Hilbert Transform of Intrinsic Mode Functions (AHTIMF) and Instantaneous Phase of the Hilbert Transform of IMFs (IPHTIMF) ; \textbf{(4) Spatial topology domain:} Functional Connectivity (FC), Differential Asymmetry (DASM), and Rational Asymmetry (RASM) \cite{jenke2014taffc}. Specific information regarding the extracted features is shown in Table~\ref{tab:dim_fea}.


\begin{table}[!t]
\caption{Comparisons between three EEG datasets.}
\centering
\renewcommand{\tabularxcolumn}[1]{m{#1}}   
\begin{tabularx}{\columnwidth}{c c c >{\centering\arraybackslash}X}
 \hline \hline
   Database     & DREAMER  & DEAP  & AMIGOS      \\ \hline
   Channel no.  & 14       & 32    & 14    \\
   Subject no.  & 23       & 32    & 40   \& 37     \\
   Video no.    & 18       & 40    & 16   \& 4     \\
   Sample no.   & 414      & 1280  & 755    \\
   Stimulus      & film clips     & music videos  & emotional videos  \\ \hline \hline
\end{tabularx}
\label{tab:comp_datasets}
\end{table}

\begin{table}[!t]
\caption{The dimensions of all EEG feature types extracted from DEAP, DREAMER, and AMIGOS.}
\begin{center}
{
\begin{tabular}{cccccc}
 \hline \hline
 \multicolumn{1}{c}{\multirow{1}*{Source domain}}  &Feature type                &DREAMER  & DEAP &AMIGOS                   \\ \hline
 \multicolumn{1}{c}{\multirow{5}*{Time}} &NSI                         & 14  & 32     &14                       \\
  &HOC                         & 140 & 320  &140                        \\
  &SPE                         & 14  & 32   &14                      \\
  &SHE                         & 14  & 32   &14                      \\
  &C0              & 14  & 32   &14                    \\ \hline
 \multicolumn{1}{c}{\multirow{3}*{Frequency}} &$AP$                          & 70  & 160         &70                 \\
  &$AP_{\beta}/AP_{\theta}$    & 14  & 32   &14                      \\
  &DE                          & 70  & 160  &70                      \\ \hline
 \multicolumn{1}{c}{\multirow{2}*{Time-frequency}} &AHTIMF                      & 70  & 160     &70                  \\
  &IPHTIMF                     & 70  & 160  &70                      \\ \hline
 \multicolumn{1}{c}{\multirow{3}*{Spatial}}  &DASM                        & 35  & 70      &35                  \\
  &RASM                        & 35  & 70   &35                      \\
  &FC                          & 91  & 496  &91                      \\ \hline
  Total    &                   & 651 & 1756 &651                      \\
 \hline \hline
\end{tabular}}
\end{center}
\label{tab:dim_fea}
\end{table}

\subsection{Comparison methods}

To comprehensively evaluate the performance of the proposed method on the multi-dimensional affective computing task, a rigorous comparative analysis is conducted against fourteen state-of-the-art feature selection techniques. The benchmarked methods include: \textbf{(1) EEG feature selection methods:} FSOR \cite{xu2020fsorer} and GRMOR \cite{GRMOR2021taffc}; \textbf{(2) Multi-label feature selection methods:} PMU \cite{lee2013feature}, FIMF \cite{lee2015fast}, SCLS \cite{lee2017scls},  MDFS \cite{zhang2019manifold}, GRRO \cite{zhang2020multi}, MGFS \cite{hashemi2020mgfs}, SCMFS \cite{hu2020multi}, and MFS\_MCDM \cite{hashemi2020mfs}; \textbf{(3) Incomplete multi-label feature selection methods:}  MLMLFS \cite{zhu2018multi}, FSMML \cite{yin2024feature}, WSEL \cite{xu2024wsel}, and WFDP\cite{dai2025multi}.



\subsection{Experimental setup}
EEG recordings underwent binary stratification according to participants' self-evaluated affective dimensions, where a threshold of 5 was applied for DEAP and AMIGOS, and a threshold of 3 was adopted for DREAMER (since DEAP and AMIGOS use a 9-point scale while DREAMER uses a 5-point scale, the median value is chosen as the threshold). Multi-Label K-Nearest Neighbors (ML-KNN) is employed as the foundational classifier, configuring neighborhood size k = 10 and smoothing parameter s = 1. The participant cohort was partitioned through randomized allocation, with 70 \% of subjects designated for training and the remaining 30 \% reserved for testing. A cross-subject experimental framework was adopted. To address inter-individual variability and ensure statistical robustness, 50 independent experimental iterations were executed. The final performance metric for affective state quantification was derived from the ensemble average of these trials. The experiment employed six distinct assessment criteria to quantify the efficacy of affective state computing. These include two label-based metrics, Macro-F1 (MA) and Micro-F1 (MI), and four example-based metrics: Average Precision (AP), Coverage (CV), Ranking Loss (RL), and Hamming Loss (HL). Further details on these evaluation measures are available in \cite{zhang2019manifold}.

This study employed a specific strategy \cite{liu2018late} to simulate an incomplete multi-label scenario by randomly removing labels proportionally on each emotion dimension to generate missing data. The label missing ratio is set from 10 \% to 50 \% with a step size of 10 \%. Following prior study~\cite{GRMOR2021taffc}, which demonstrated that around 10 \% of EEG features are sufficient for effective affective computing, we therefore set the feature selection ratio to 10 \% when applying ADSEL. The tuning range for the trade-off parameters ( $\alpha$, $\lambda$, $\beta$, $\mu$, and $\delta$) is set to $10^{-3}$ to $10^{3}$ with a tuning step size of $10^{1}$.

\begin{figure*}[!t]
\centering
\subfloat[HL\label{HL_eva_deap}]{\includegraphics[width=0.23\textwidth]{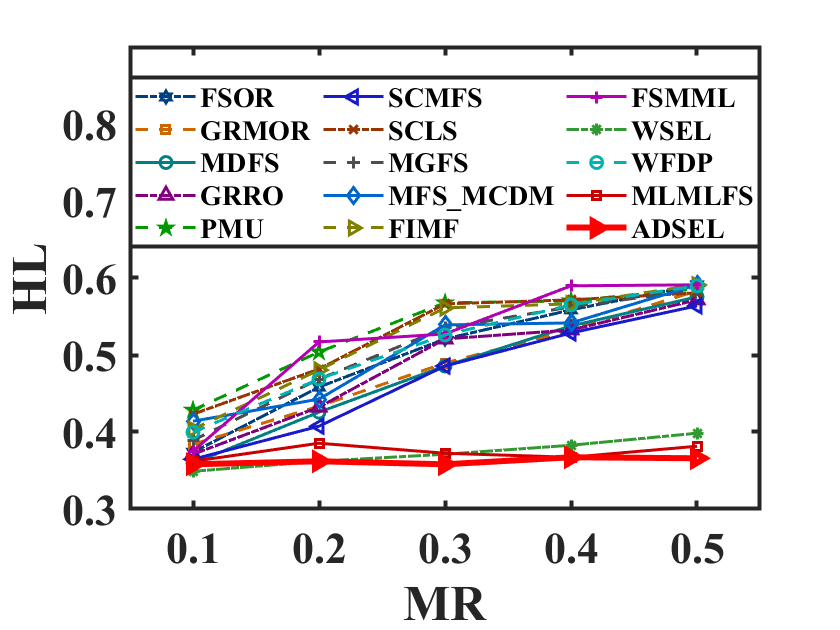}}
\subfloat[RL\label{CV_eva_deap}]{\includegraphics[width=0.23\textwidth]{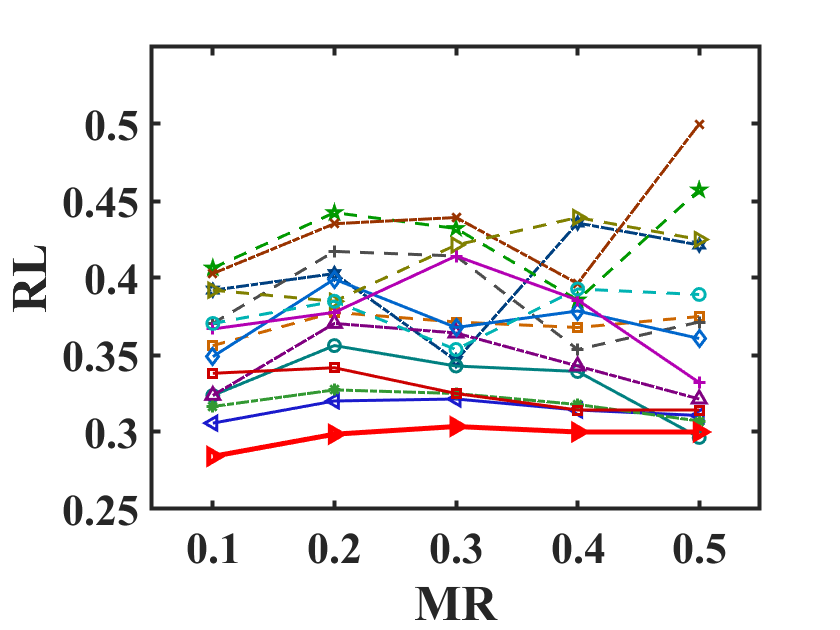}}
\subfloat[CV\label{RL_eva_deap}]{\includegraphics[width=0.23\textwidth]{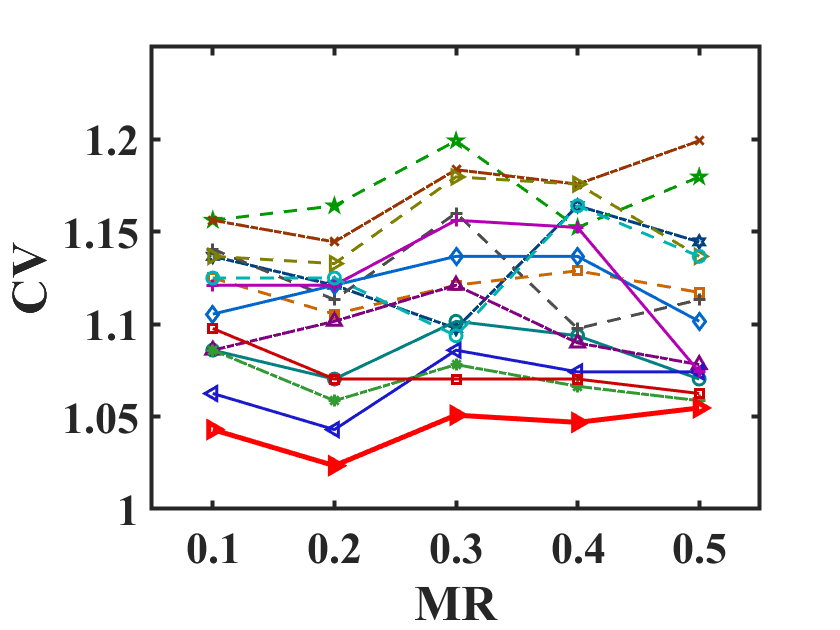}}

\vspace{-0.2cm}
\subfloat[AP\label{AP_eva_deap}]{\includegraphics[width=0.23\textwidth]{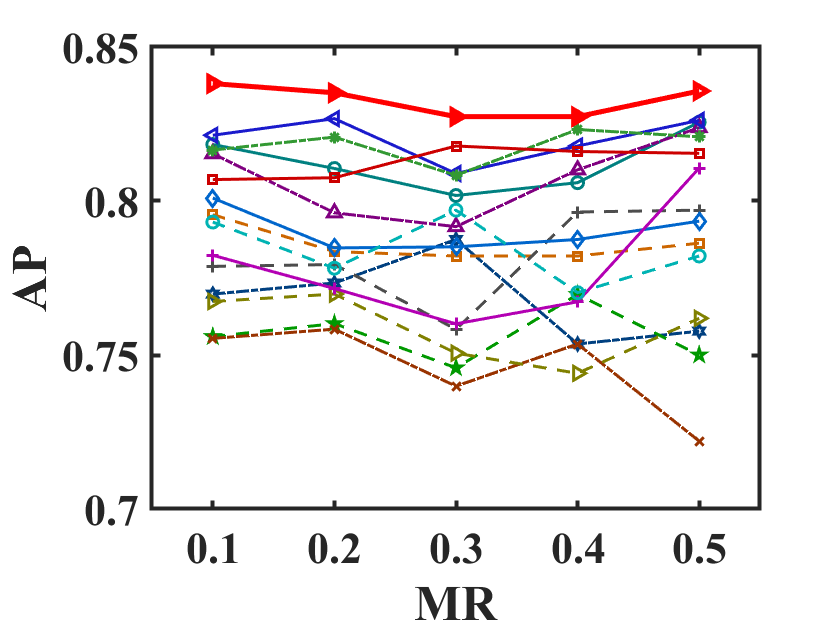}}
\subfloat[MA\label{MA_eva_deap}]{\includegraphics[width=0.23\textwidth]{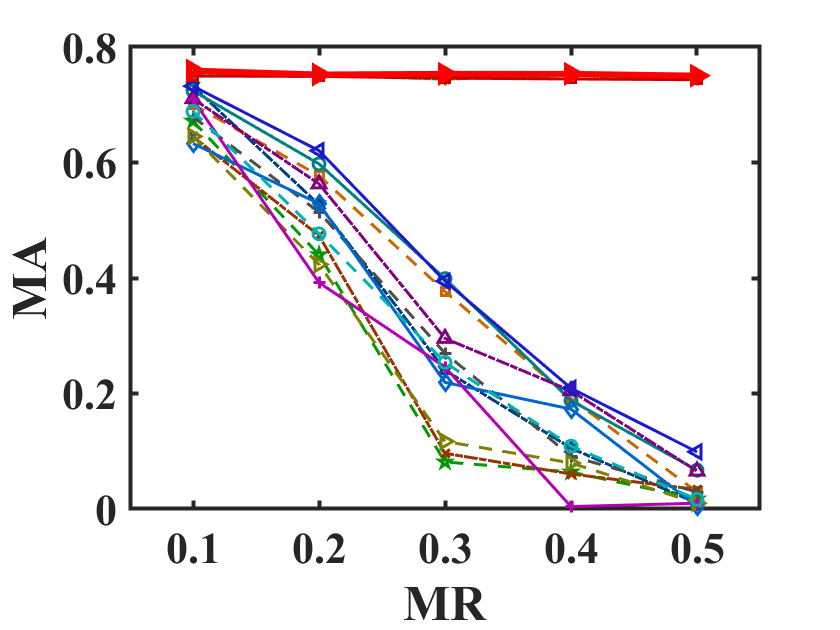}}
\subfloat[MI\label{MI_eva_deap}]{\includegraphics[width=0.23\textwidth]{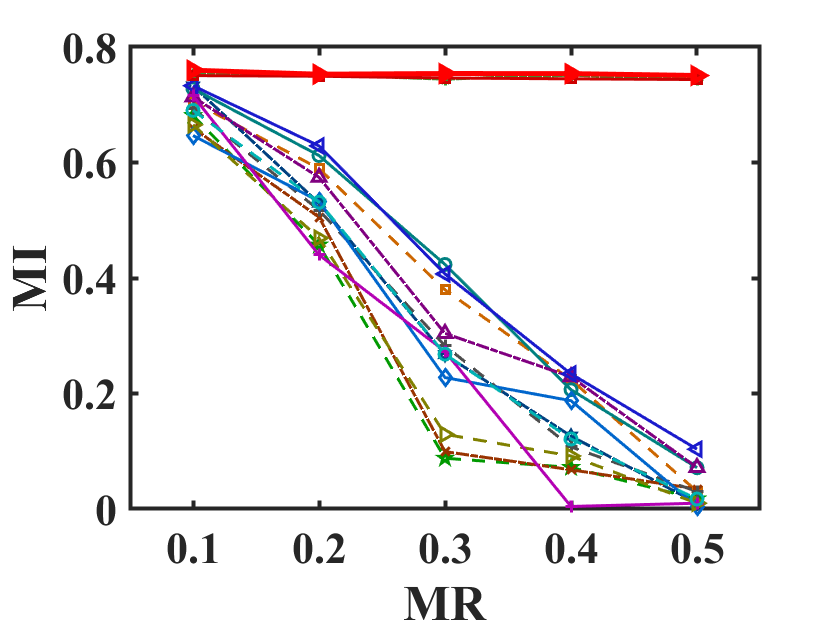}}
\caption{Multi-dimensional emotion recognition performance of various label missing ratios (MR) on DEAP.}
\label{Results_index_deap}
\end{figure*}

\begin{figure*}[!t]
\centering
\subfloat[HL\label{HL_eva_dreamer}]{\includegraphics[width=0.23\textwidth]{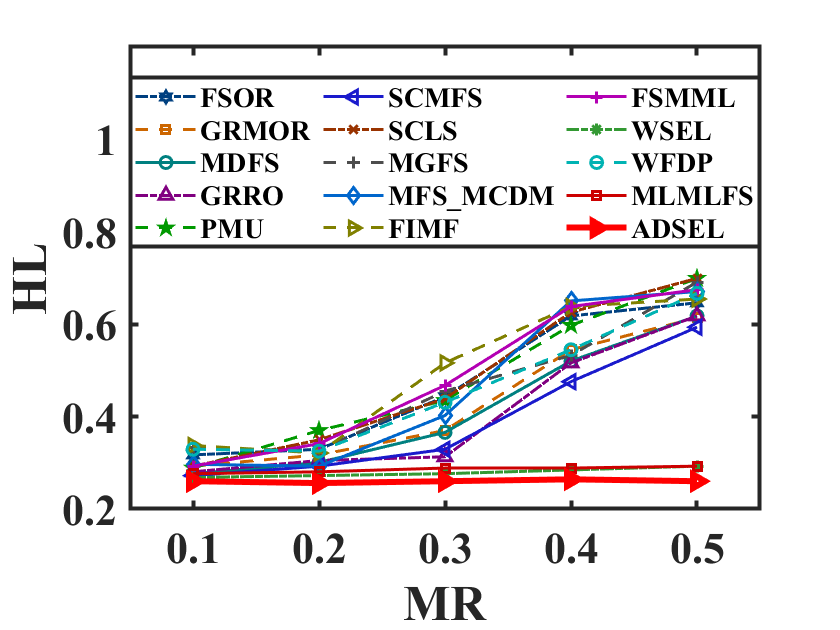}}
\subfloat[RL\label{CV_eva_dreamer}]{\includegraphics[width=0.23\textwidth]{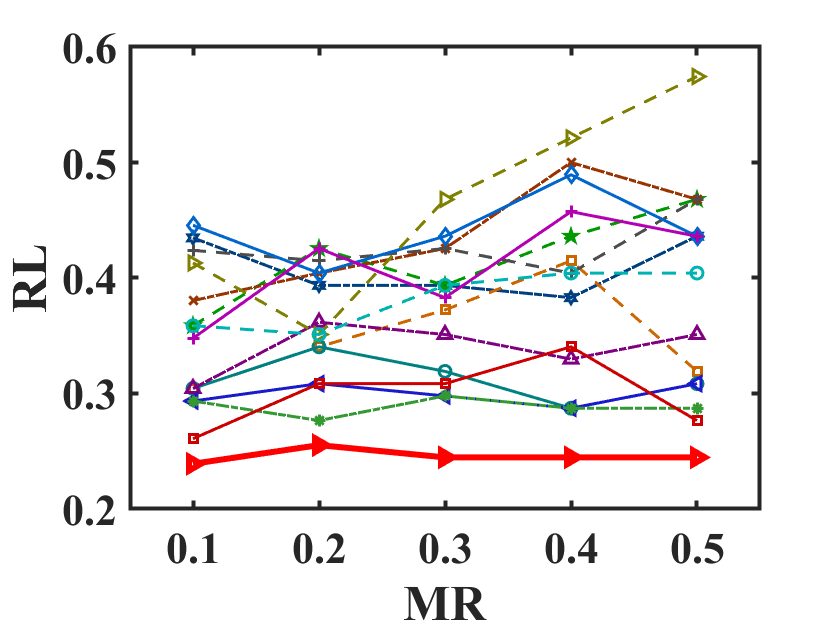}}
\subfloat[CV\label{RL_eva_dreamer}]{\includegraphics[width=0.23\textwidth]{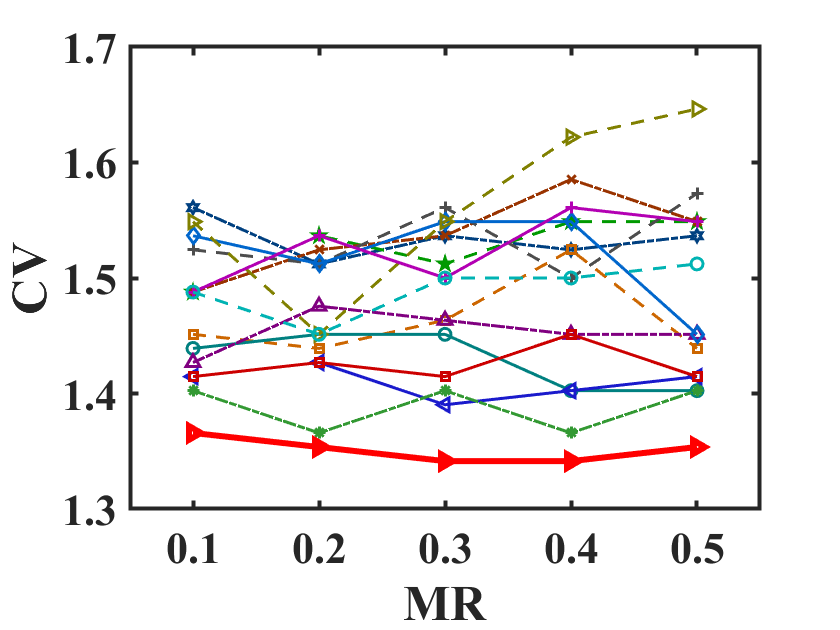}}

\vspace{-0.2cm}
\subfloat[AP\label{AP_eva_dreamer}]{\includegraphics[width=0.23\textwidth]{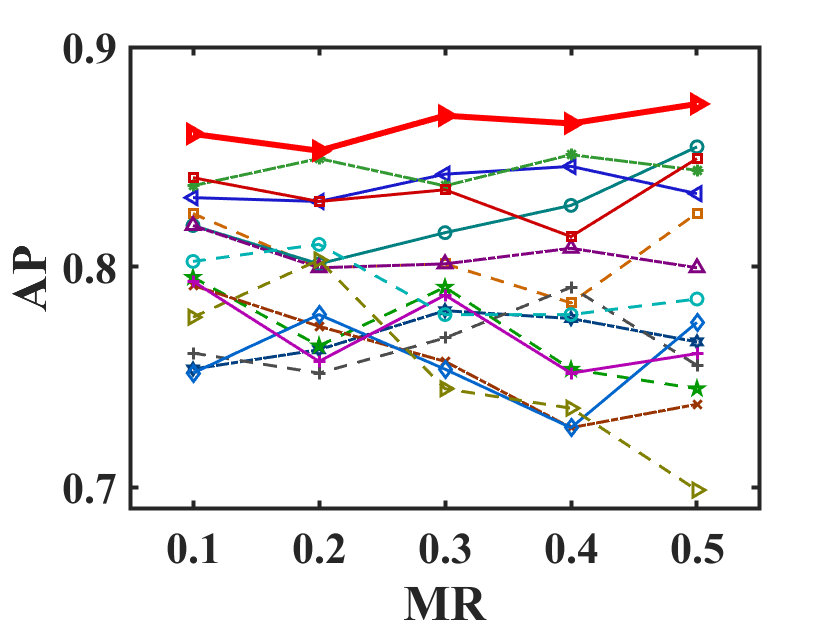}}
\subfloat[MA\label{MA_eva_dreamer}]{\includegraphics[width=0.23\textwidth]{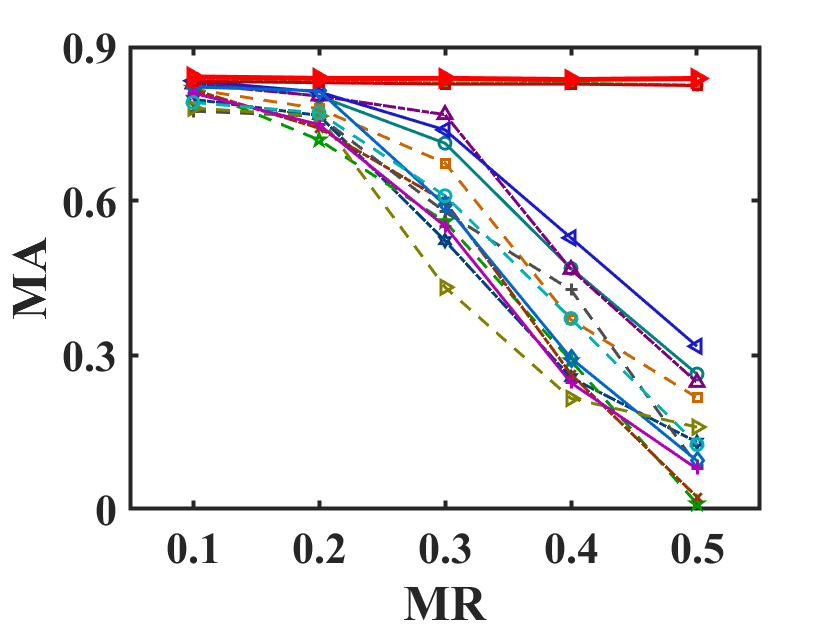}}
\subfloat[MI\label{MI_eva_dreamer}]{\includegraphics[width=0.23\textwidth]{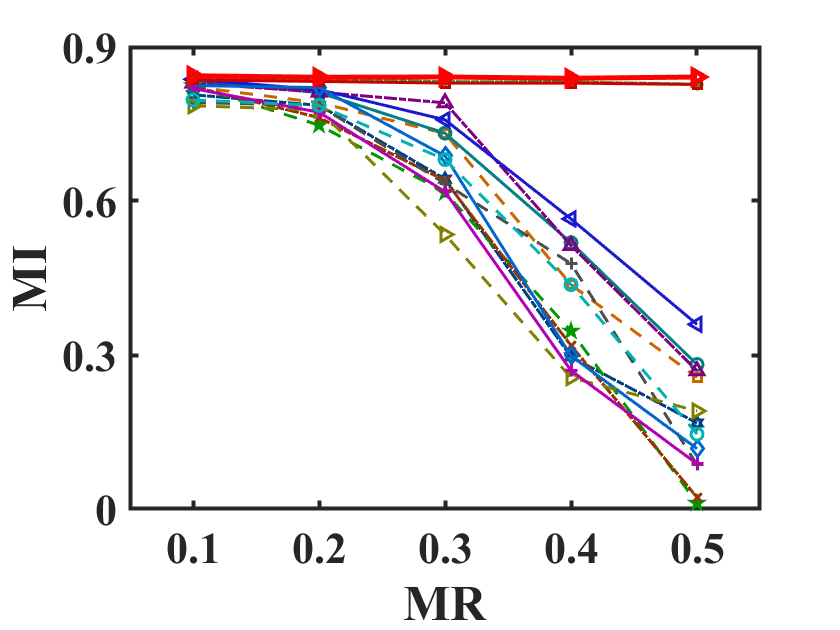}}
\caption{Multi-dimensional emotion recognition performance of various label MR on DREAMER.}
\label{Results_index_dreamer}
\end{figure*}

\begin{figure*}[!t]
\centering
\subfloat[HL\label{HL_eva_amigos}]{\includegraphics[width=0.23\textwidth]{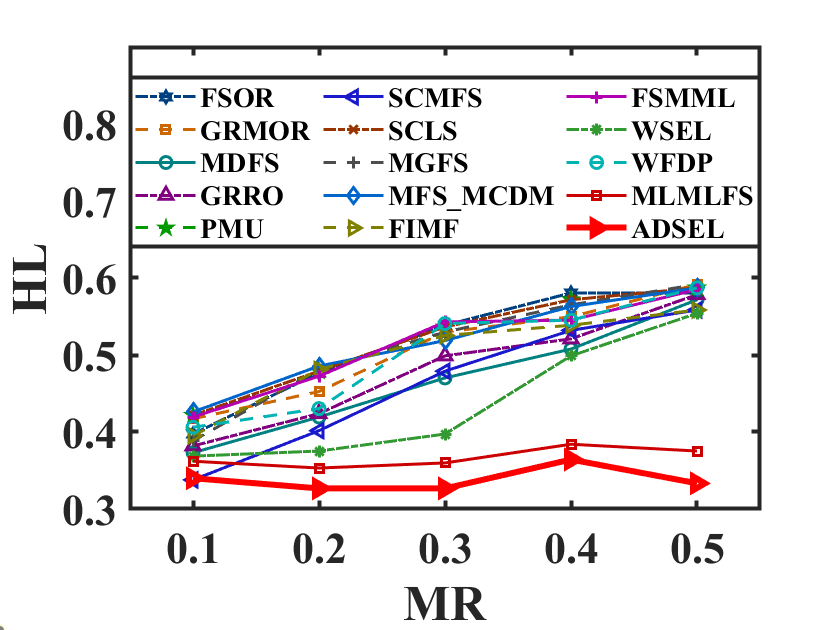}}
\subfloat[RL\label{CV_eva_amigos}]{\includegraphics[width=0.23\textwidth]{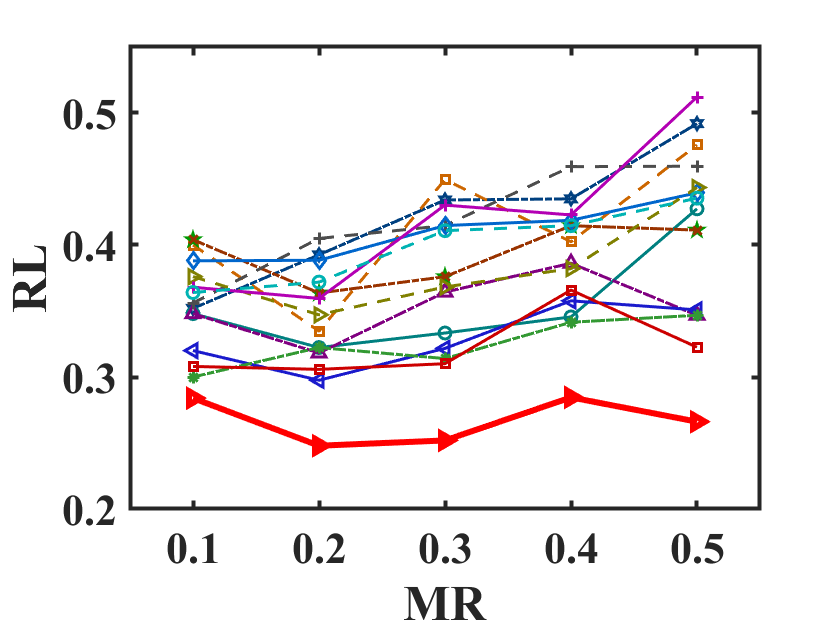}}
\subfloat[CV\label{RL_eva_amigos}]{\includegraphics[width=0.23\textwidth]{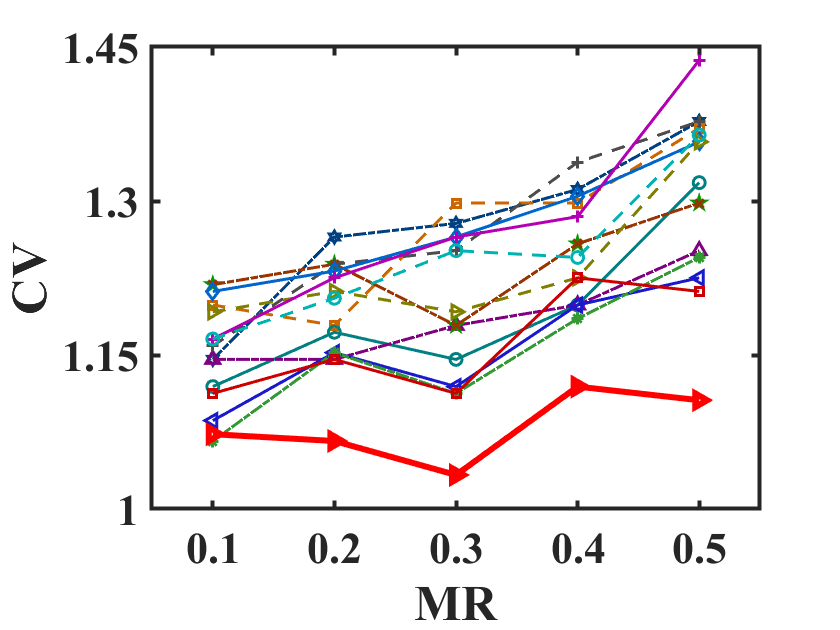}}

\vspace{-0.2cm}
\subfloat[AP\label{AP_eva_amigos}]{\includegraphics[width=0.23\textwidth]{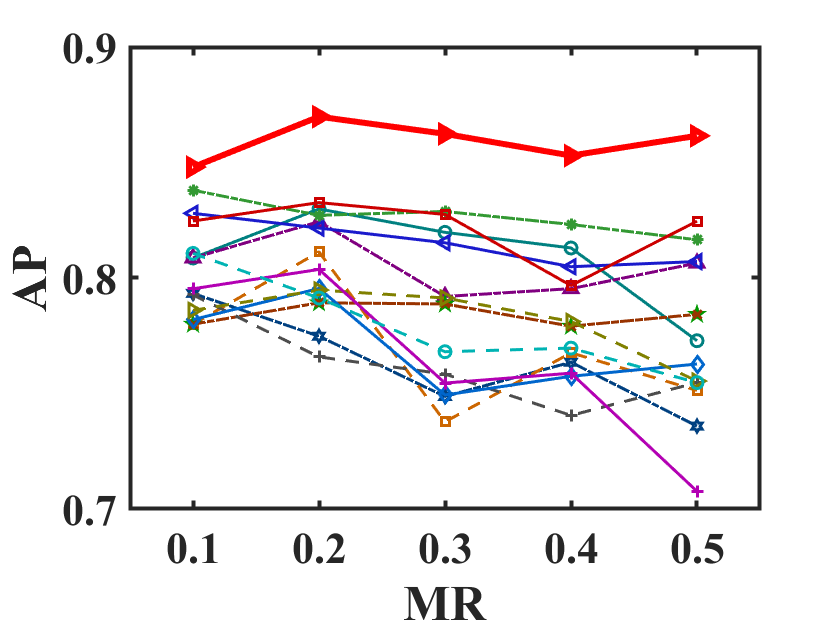}}
\subfloat[MA\label{MA_eva_amigos}]{\includegraphics[width=0.23\textwidth]{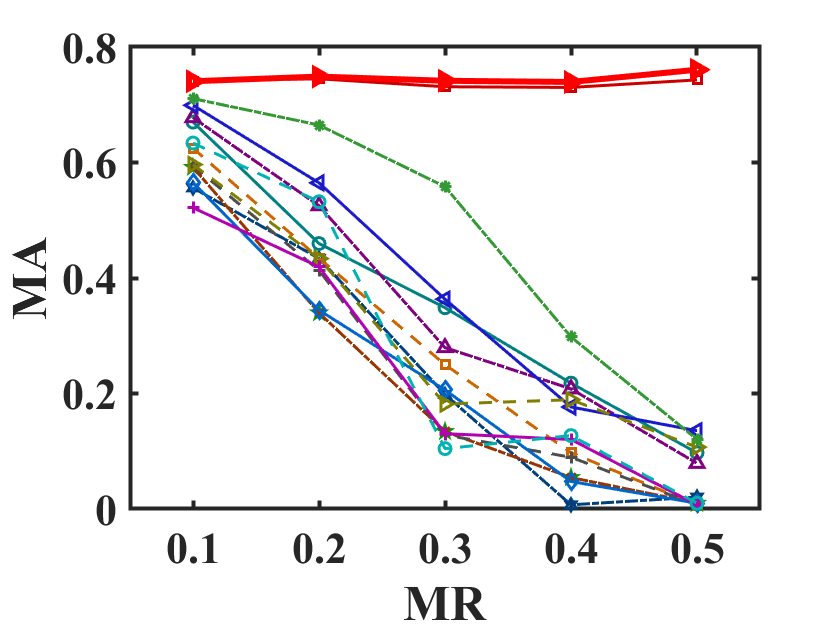}}
\subfloat[MI\label{MI_eva_amigos}]{\includegraphics[width=0.23\textwidth]{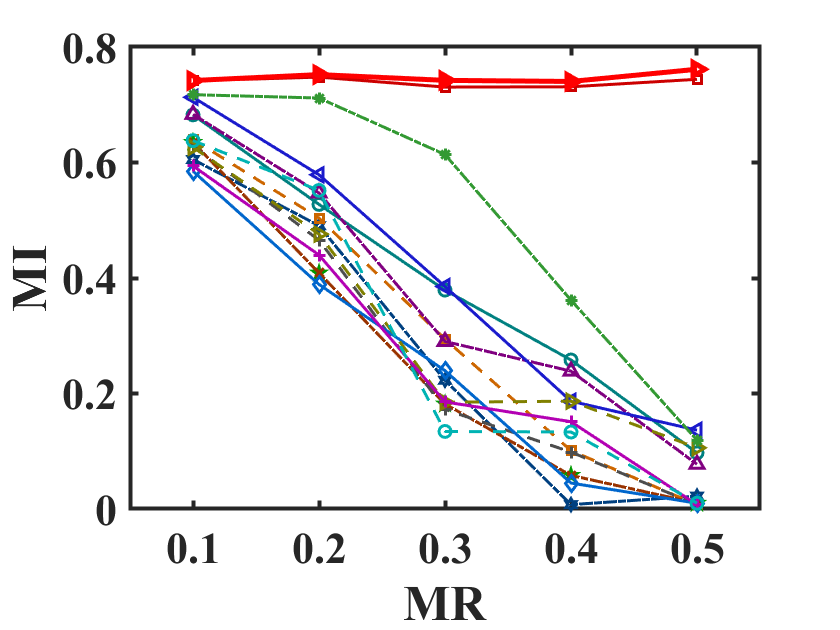}}
\caption{Multi-dimensional emotion recognition performance of various label MR on AMIGOS.}
\label{Results_index_amigos}
\end{figure*}

\section{Experimental Results and Discussion}\label{Experimental Results and Discussion}

\subsection{Emotion recognition performance with incomplete labels}
Figs.~\ref{Results_index_deap}-\ref{Results_index_amigos} present the emotion recognition comparison results on the DEAP, DREAMER, and AMIGOS datasets. The horizontal axis represents the missing ratios of multi-dimensional emotion labels, and the vertical axis corresponds to the evaluation results of various performance metrics. The experimental results of the ADSEL method are marked with red lines in Figs.~\ref{Results_index_deap}-\ref{Results_index_amigos}. As illustrated in Figs.~\ref{Results_index_deap}-\ref{Results_index_amigos} (a-c), the multi-dimensional emotion recognition capability consistently increases with lower values. Conversely, as shown in Figs.~\ref{Results_index_deap}-\ref{Results_index_amigos} (d-f) the multi-dimensional emotion recognition capability rises with higher values. Compared to the other fourteen comparison methods, ADSEL has the optimal performance under different missing ratios. As illustrated in Fig. 2(a–c) and Fig. 3(a–c), the multi-dimensional emotion recognition capability consistently increases with lower values. Conversely, as shown in Fig. 2(d–f) and Fig. 3(d–f), the same capability rises with higher values.
Synthesizing the evidence from Figs.~\ref{Results_index_deap}-\ref{Results_index_amigos}, it is unequivocally demonstrated that the EEG feature subset selected by the proposed ADSEL method consistently achieves the most competitive performance in the incomplete multi-dimensional emotion recognition task.

\begin{figure*}[!t]
\centering
\subfloat[HL]{\label{nemenyi_hl}\includegraphics[width=0.3\textwidth]{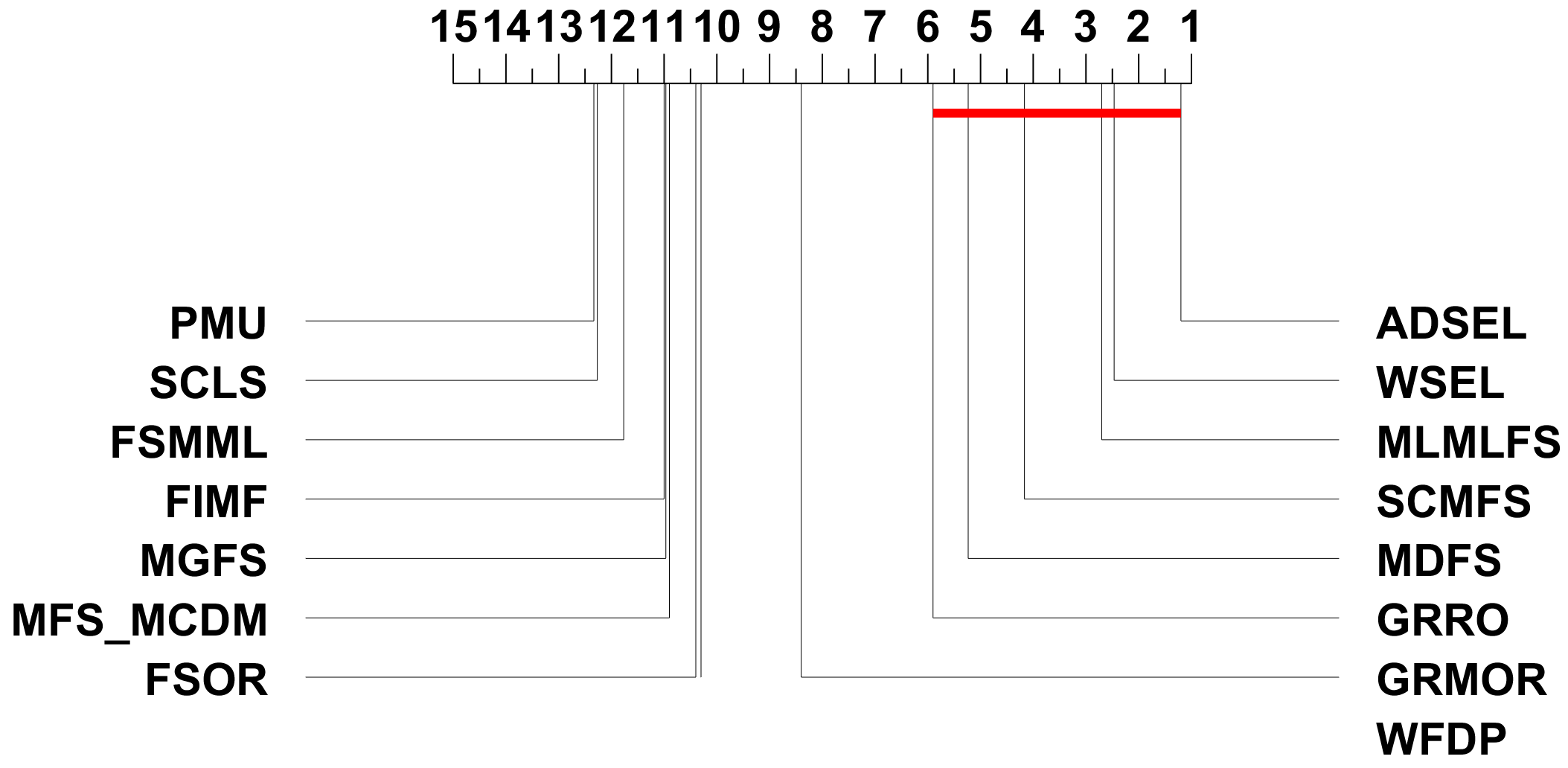}}
\hspace{0.5cm}
\subfloat[AP]{\label{nemenyi_ap}\includegraphics[width=0.3\textwidth]{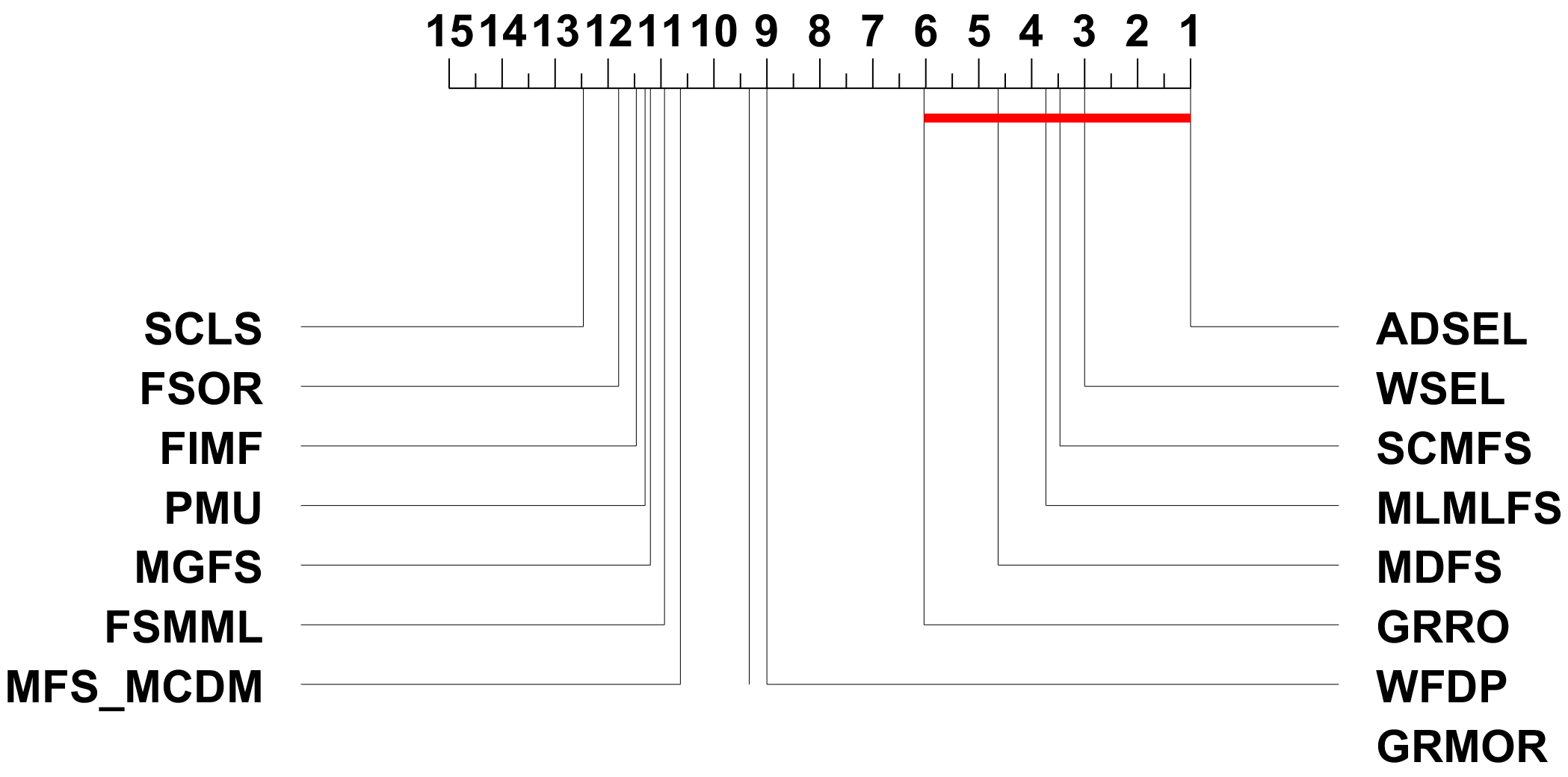}}
\hspace{0.5cm}
\subfloat[MA]{\label{nemenyi_ma}\includegraphics[width=0.3\textwidth]{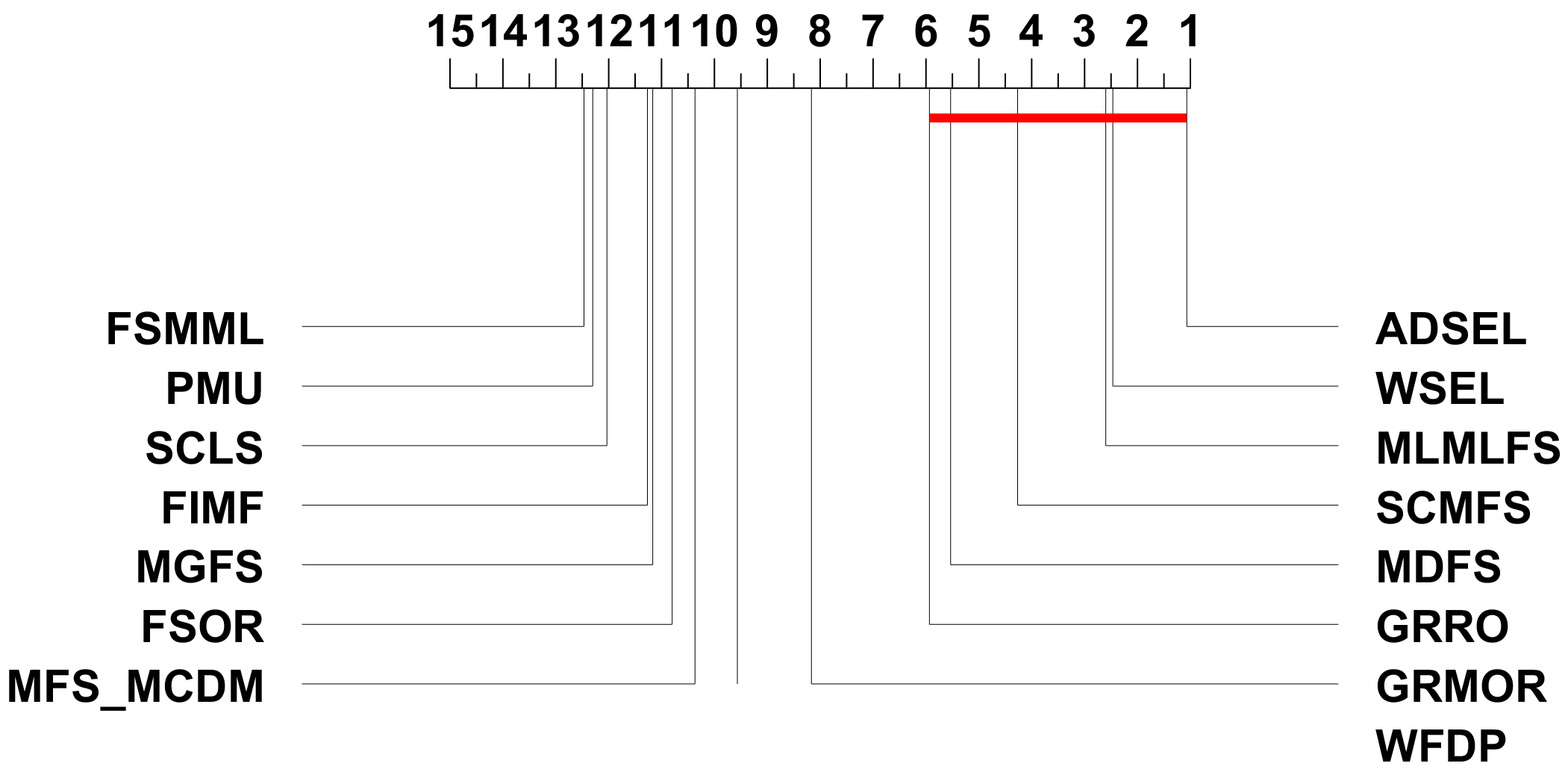}}
\hspace{0.5cm}
\caption{The Nemenyi test results (CD=12.3858 , significance level $\alpha=0.05$ ): (a) HL; (b) AP; (c) MA.}
\label{Nemenyi_dreamer}
\end{figure*}

To rigorously evaluate the comparative performance of the fifteen feature selection techniques, the Nemenyi test was performed, and the resulting critical difference diagram along with detailed comparison outcomes are shown in Fig.~\ref{Nemenyi_dreamer}. Across all evaluation metrics, the proposed ADSEL method consistently achieves the lowest mean rank. Moreover, the pairwise comparisons confirm the statistical significance of its superiority over each of the other fourteen techniques.

\subsection{Label reconstruction results}
\begin{figure}[!t]
\centering
\subfloat[DEAP\label{deap}]{\includegraphics[width=0.16\textwidth]{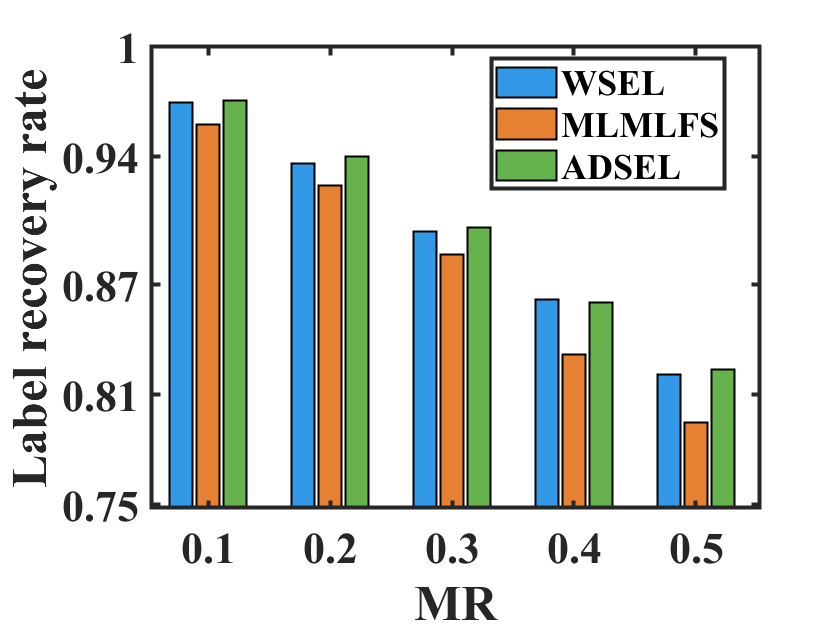}}
\subfloat[DREAMER\label{dreamer}]{\includegraphics[width=0.16\textwidth]{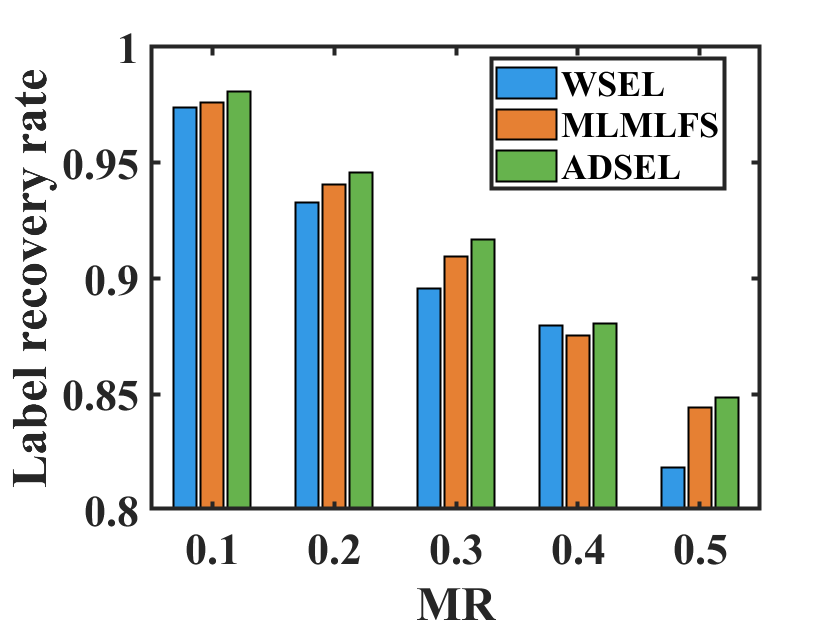}}
\subfloat[AMIGOS\label{amigos}]{\includegraphics[width=0.16\textwidth]{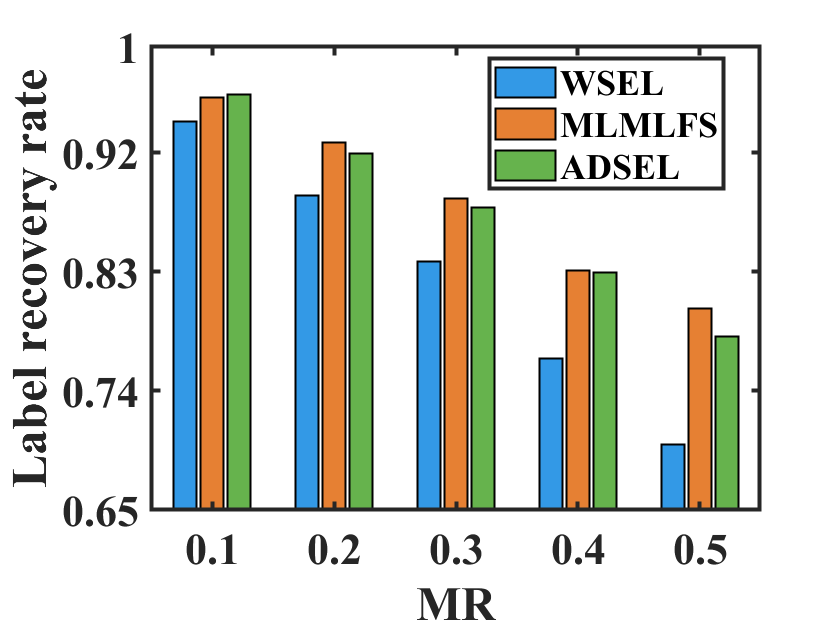}}
\vspace{-0.2cm}
\caption{Label recovery ratios on the three data sets.}
 \label{label_recovery_ratio}
\end{figure}

The label recovery capability under incomplete multi-label settings is further assessed, with results detailed in Fig.~\ref{label_recovery_ratio}, which shows average label recovery accuracy across three datasets under various missing rates. ADSEL yields the highest average accuracy on DREAMER and DEAP. On AMIGOS, it ranks second but trails the best-performing by a marginal margin, suggesting competitive robustness. The favorable label recovery performance is attributed to the adaptive dual self-expression learning mechanism: unlike methods that model sample-wise similarity or label correlation separately, ADSEL introduces a joint regularization term that couples both perspectives. This coupling helps recover missing labels by leveraging complementary information, especially when one view suffers from high noise. Overall, these results indicate that the proposed mechanism offers a viable solution for robust label imputation under the evaluated missing-label scenarios.

\subsection{Robustness analysis of feature selection}
\begin{figure*}[!t]
\centering
\subfloat[DEAP\label{deap}]{\includegraphics[width=0.33\textwidth]{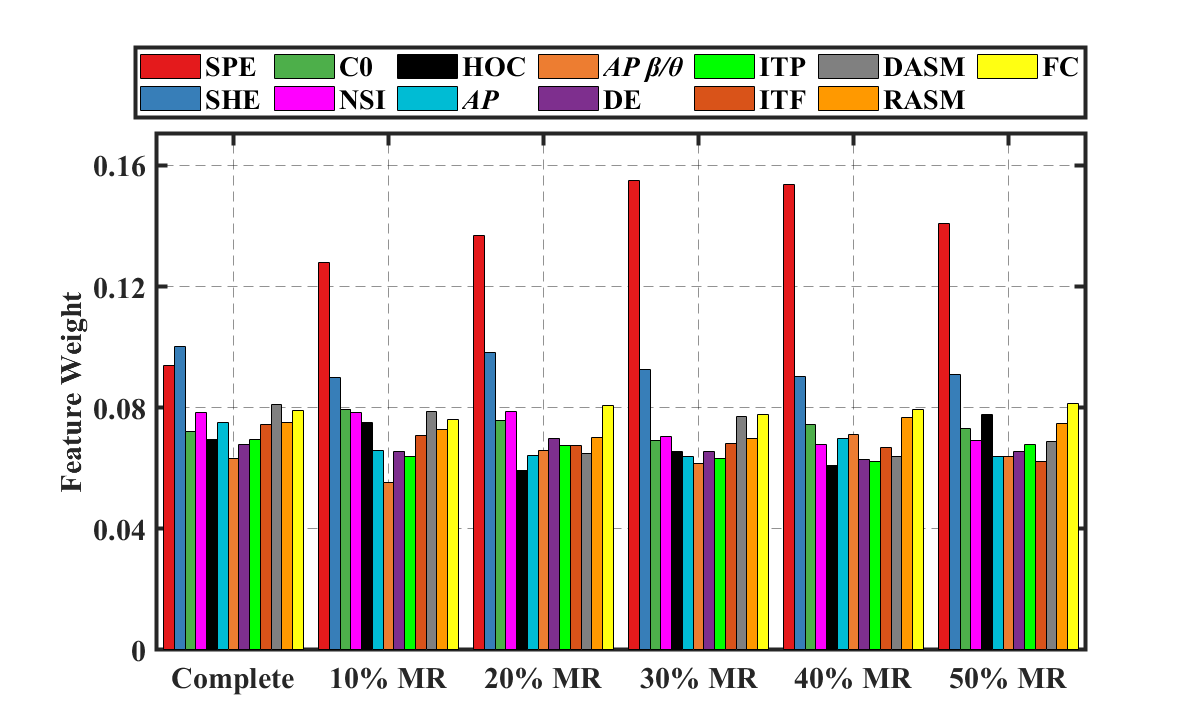}}
\subfloat[DREAMER\label{dreamer}]{\includegraphics[width=0.33\textwidth]{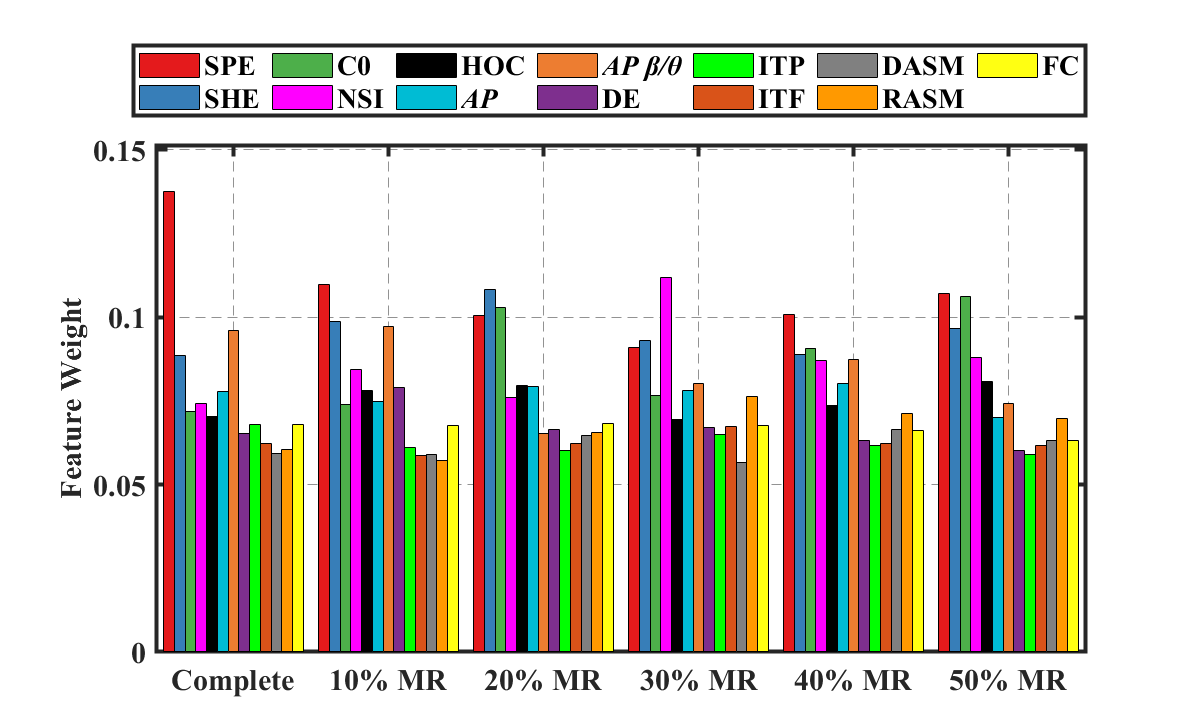}}
\subfloat[AMIGOS\label{amigos}]{\includegraphics[width=0.33\textwidth]{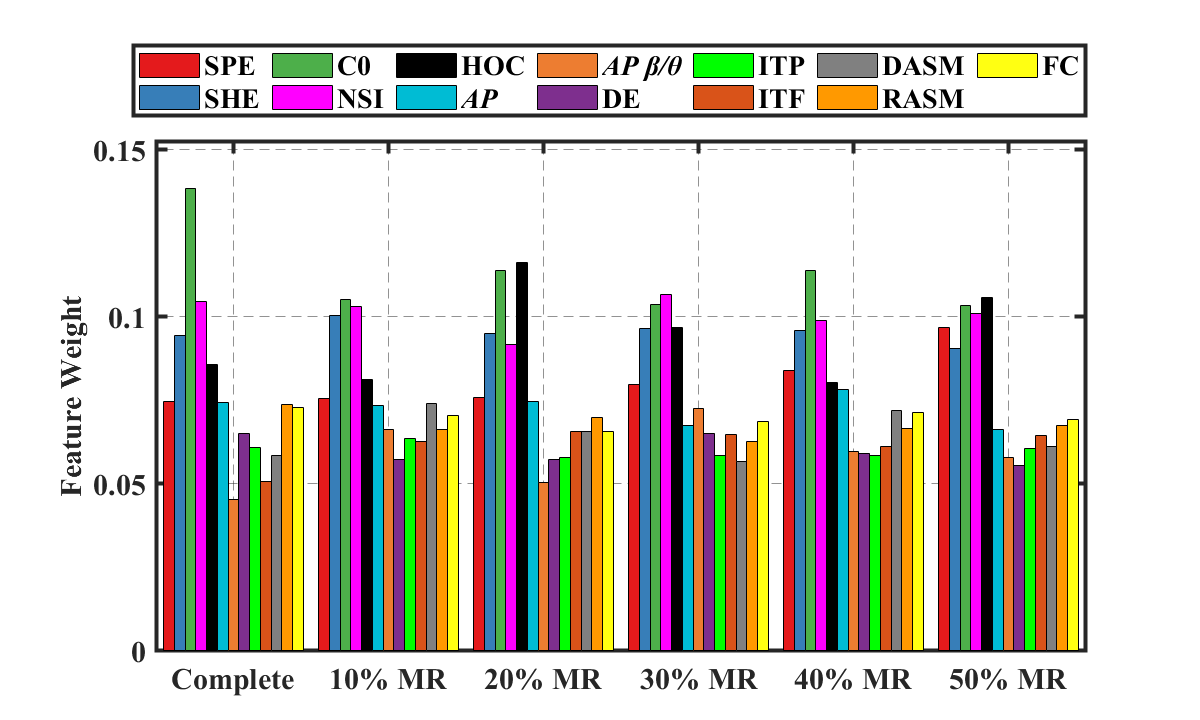}}
\caption{Feature weight analysis under distinct label missing ratios.}
\label{label_feature_weight}
\end{figure*}

To comprehensively assess the stability of the proposed feature selection method, separate analyses were performed on feature weights. The feature weights are computed as the normalized reciprocal of the average scores obtained from the ADSEL method, reflecting each feature’s contribution to multi-dimensional emotion recognition. Fig.~\ref{label_feature_weight} illustrates the distribution of feature weights across three datasets under varying label missing rates. Notably, features such as SHE and SPE consistently exhibit high weights. Their stability under incomplete label conditions provides strong evidence that ADESL reliably identifies discriminative features even when the proportion of labeled samples is severely limited, thereby demonstrating its robustness in label-scarce scenarios.

\subsection{Parameter sensitivity}
\begin{figure}[!t]
\centering
\subfloat[$\alpha$\label{alpha}]{\includegraphics[width=0.15\textwidth]{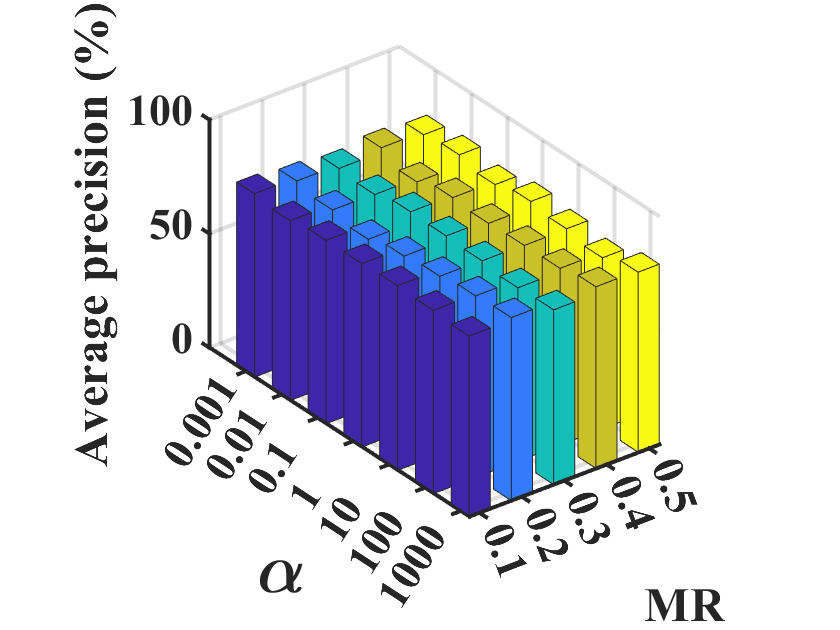}}
\subfloat[$\beta$\label{beta}]{\includegraphics[width=0.15\textwidth]{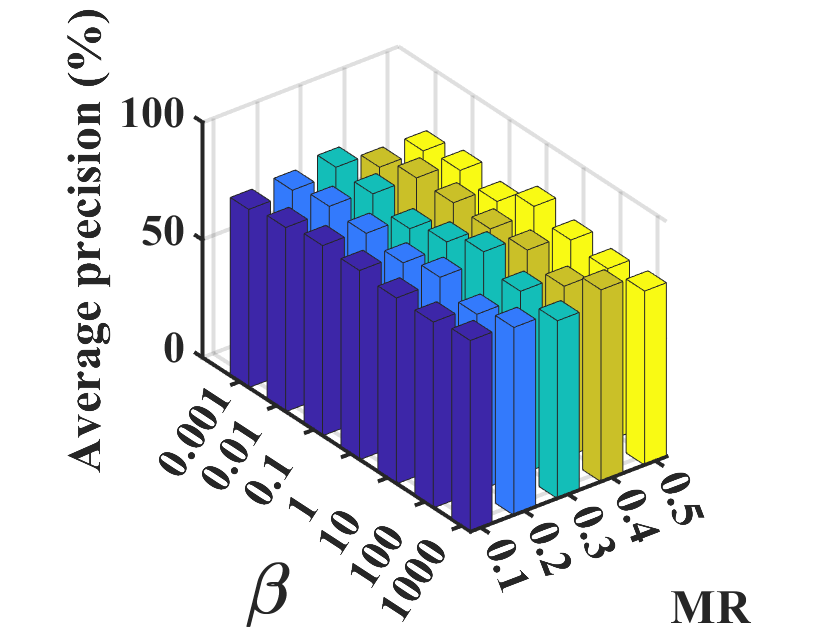}}
\subfloat[$\lambda$\label{lambda}]{\includegraphics[width=0.15\textwidth]{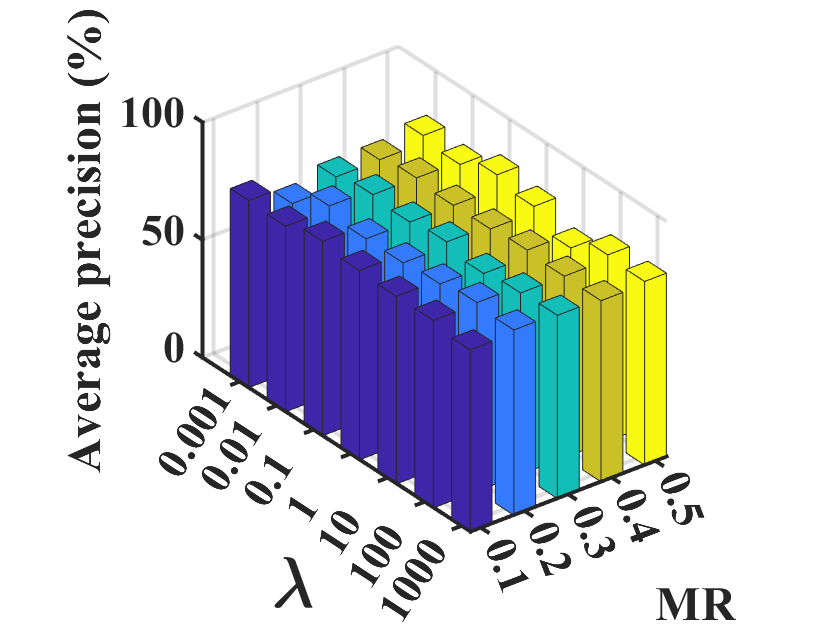}}

\vspace{-0.2cm}
\subfloat[$\mu$\label{mu}]{\includegraphics[width=0.15\textwidth]{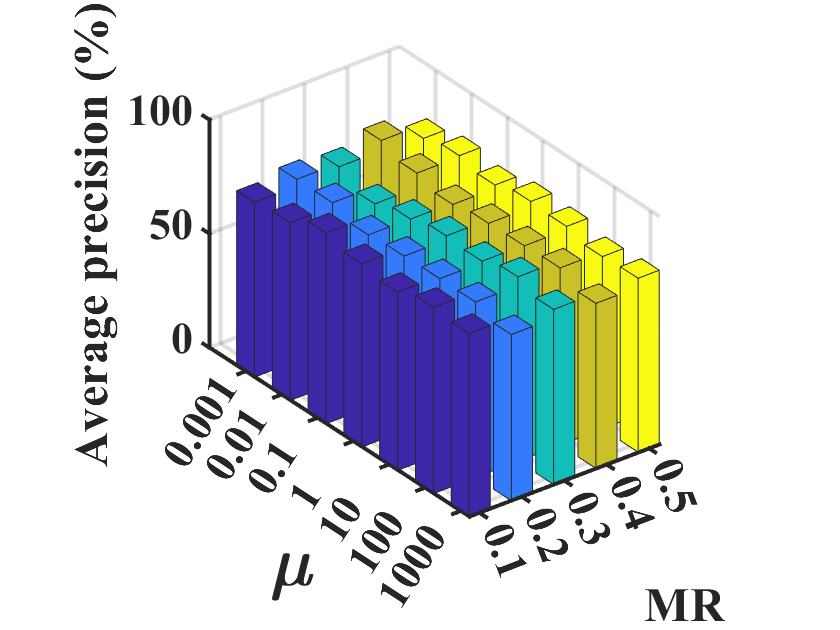}}
\subfloat[$\delta$\label{delta}]{\includegraphics[width=0.15\textwidth]{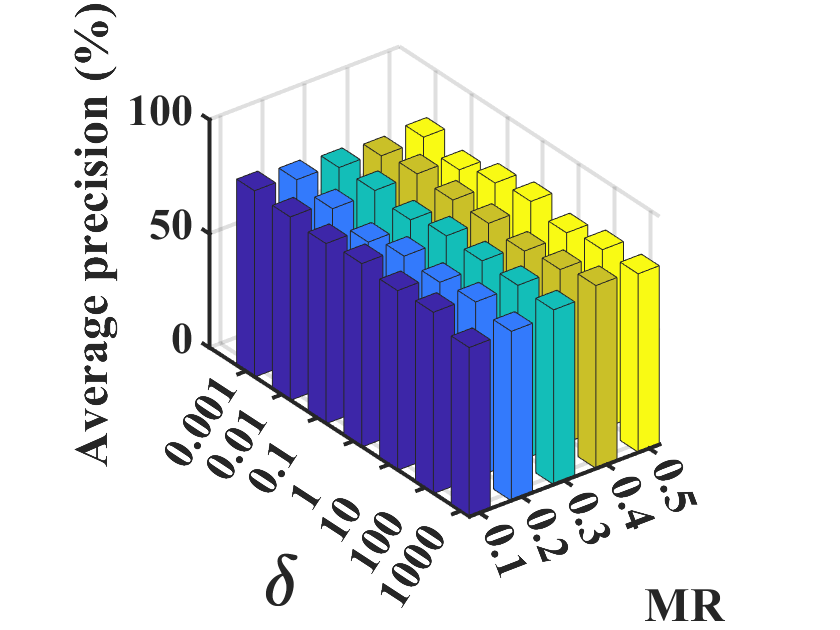}}
\caption{The parameter sensitivity on the DREAMER dataset.}
\label{parameter_sensitivity_dreamer}
\end{figure}

In order to explore the sensitivity of ADSEL to the five regularization parameters, an experiment was carried out to evaluate the impact of them and report the arising performance differences as shown in Fig.~\ref{parameter_sensitivity_dreamer} (on the DREAMER dataset). During the experiment, each parameter was varied individually while the others were set to 10. When parameters change, the fluctuations in performance metrics are small, indicating that ADSEL is not very sensitive to parameter changes.

\begin{table}[!t]
\caption{Ablation study results in terms of AP, where w/o denotes the baseline (without the component), ALL is adaptive label learning, GFRL is global feature redundancy learning, and GMR is the graph-based manifold regularizer.}
\begin{center}
{
\begin{tabular}{cccc}
\hline\hline
\multicolumn{1}{l}{\multirow{2}*{Conditions}}         & \multicolumn{3}{c}{\multirow{1}*{Average precision}} \\  \cline{2-4}
                          & DEAP           & DREAMER     &AMIGOS           \\\hline
w/o Matrix $\mathbf{Q}$ of ALL     & 0.83           & 0.84        & 0.84     \\
w/o Matrix $\mathbf{U}$ of ALL     & 0.82           & 0.84        & 0.83     \\
w/o ALL                   & 0.75           & 0.77        & 0.77     \\
w/o GFRL                  & 0.79           & 0.82        & 0.81     \\
w/o GMR                   & 0.78           & 0.81        & 0.80     \\
Our method                & 0.83           & 0.86        & 0.86     \\ \hline\hline
\end{tabular}}
\end{center}
\label{tab:acml}
\end{table}
\vspace{-0.2cm}

\begin{figure}[!t]
\centering
\subfloat[DEAP\label{conv_deap}]{\includegraphics[width=0.16\textwidth]{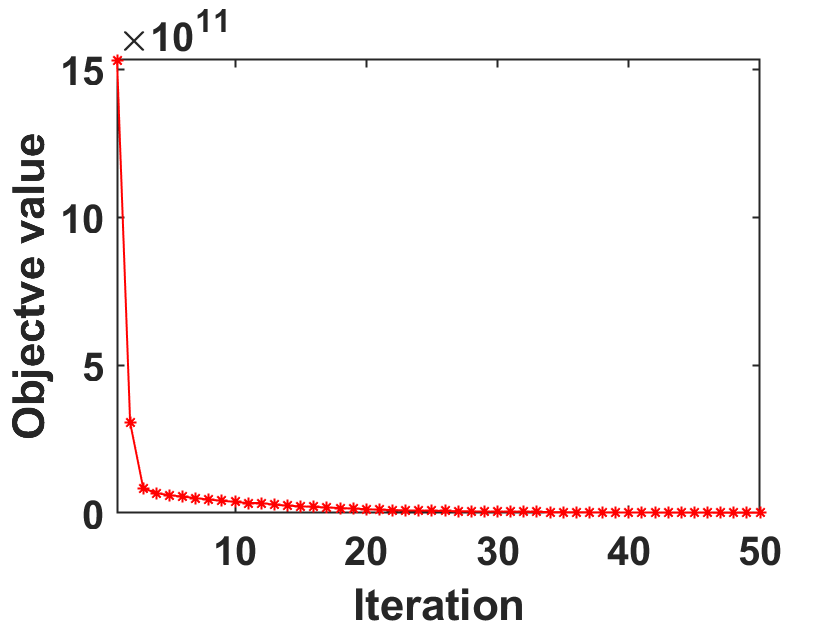}}
\subfloat[DREAMER\label{conv_dreamer}]{\includegraphics[width=0.16\textwidth]{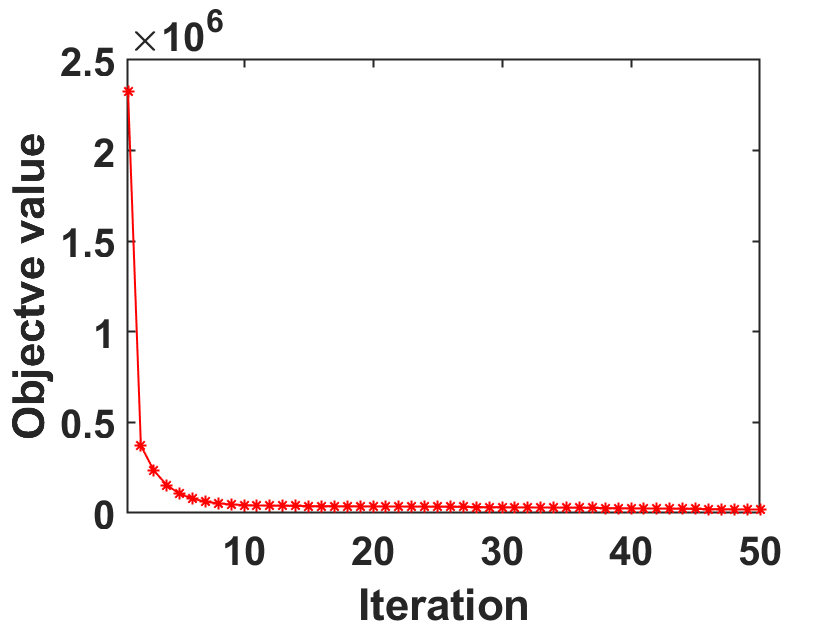}}
\subfloat[AMIGOS\label{conv_amigos}]{\includegraphics[width=0.16\textwidth]{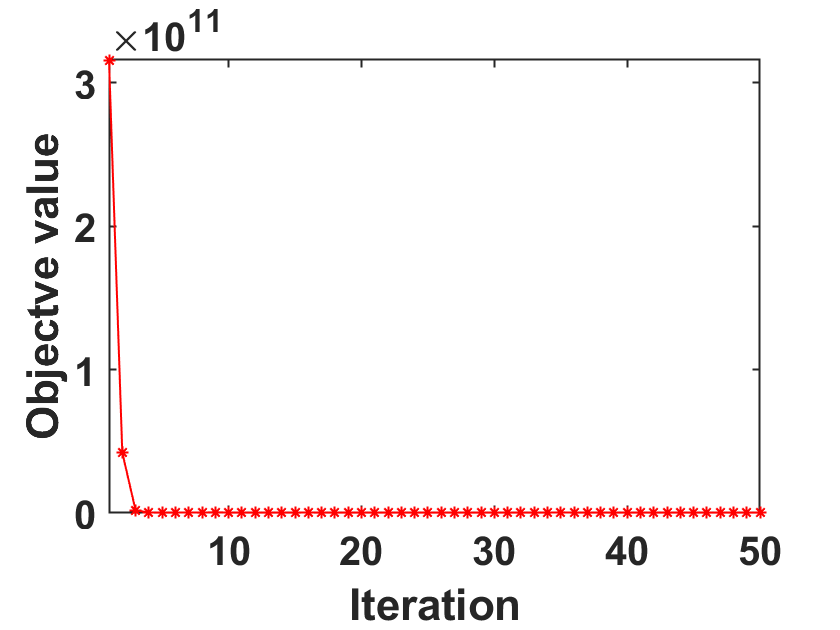}}
\caption{Convergence verification of ADSEL.}\label{conv_res}
\label{Convergence_verificationr}
\end{figure}
\vspace{-0.2cm}

\subsection{Ablation experiments}
To systematically evaluate the contributions of individual modules, we designed five ablation configurations targeting different components of the ADSEL framework. Specifically, three configurations are dedicated to dissecting the Adaptive Label Learning (ALL) module: (1) \textbf{w/o Matrix $\mathbf{Q}$}: retains only the dimension-level coefficient matrix $\mathbf{U}$ while discarding the sample-level coefficient $\mathbf{Q}$; (2) \textbf{w/o Matrix $\mathbf{U}$}: retains only the sample-level coefficient $\mathbf{Q}$ while removing the dimension-level coefficient $\mathbf{U}$; (3) \textbf{w/o ALL}: excludes the entire dual self-expression mechanism and directly utilizes the original incomplete label matrix without reconstruction. The remaining two configurations are designed to evaluate the Global Feature Redundancy Learning (GFRL) module and the Graph-based Manifold Regularization (GMR) module: (4) \textbf{w/o GFRL}: removes the Global Feature Redundancy Learning module; (5) \textbf{w/o GMR}: removes the Graph-based Manifold Regularization module.

Table~\ref{tab:acml} reports the Average Precision under a 30 \% missing-label ratio. The full ADSEL consistently outperforms all ablation configurations. Specifically, removing $\mathbf{Q}$ or $\mathbf{U}$ leads to performance degradation on DREAMER and AMIGOS, while on DEAP, the impact of removing $\mathbf{Q}$ is marginal, which still confirms that both sample-wise and dimension-wise correlations collectively contribute to robust label reconstruction, although their relative importance may vary across datasets. Furthermore, excluding GFRL or GMR also results in noticeable accuracy drops, validating their respective roles in mitigating inter-feature redundancy and preserving local manifold structures. Collectively, these ablation results substantiate the effectiveness of all three core components and further demonstrate that the dual self-expression mechanism, which jointly exploits sample-level and dimension-level dependencies within the label space, benefits from both perspectives, corroborating the core rationale of our methodological design.

\begin{table}[!t]
\caption{Computational time comparison (seconds).}
\begin{center}
{
\begin{tabular}{lccc}
\hline\hline
Methods         &  DEAP       & DREAMER   &AMIGOS\\\hline
MDFS                     & 4.08       & 1.45   &6.10        \\
GRRO                     & 32.42      & 2.01   &3.01    \\
PMU                      & 134.19     &7.54    &30.40     \\
SCMFS                    & 3.60       &0.47    &0.09    \\
WSEL                     & 3.98       &0.99    &3.66   \\ 
SCLS                     & 27.51      & 1.73   &6.59    \\
MGFS                     & 0.45       & 0.34   &0.13     \\
MFS\_MCDM                & 0.65       & 0.18   &0.12       \\
FIMF                     & 0.03       & 0.01   &0.03      \\
FSOR                     & 369.05     & 46.42  &34.59      \\
GRMOR                    & 146.22     & 27.47  &34.02        \\
FSMML                    & 12731.25   & 4336.28     &2171.37            \\
WFDP                     & 32.07      & 7.12        &7.62   \\
MLMLFS                   & 4.20       & 0.43        &0.99      \\
\textbf{ADSEL}            & \textbf{8.79}          & \textbf{0.51}       &\textbf{4.15} \\ \hline\hline
\end{tabular}}
\end{center}
\label{tab:avetime}
\end{table}

\subsection{Convergence demonstration}
Finally, the convergence speed of the proposed iterative optimization algorithm is analyzed. Fig.~\ref{conv_res} depicts the convergence behavior, plotting the objective function value versus iteration number on three datasets. All trade-off parameters ($\lambda$, $\beta$, $\alpha$, $\mu$, and  $\delta$) were uniformly set to 10. The results in Fig.~\ref{conv_res} demonstrate that the ADSEL algorithm exhibits rapid convergence, and typically stabilizing within a small number of iterations. This confirms the efficacy of our iterative optimization strategy.

\subsection{Computational cost}
The computational time requirements of all methods were also evaluated. Experiments were conducted in MATLAB (MathWorks Inc., Novi, MI, USA) on a Windows 11 x64 system equipped with an Intel Core i5-124000HQ 2.5 GHz CPU and 16.00 GB RAM. Average computational times are summarized in Table~\ref{tab:avetime}. Combining the results presented in Figs.~\ref{Results_index_deap}-\ref{Results_index_amigos} and Table~\ref{tab:avetime}, the proposed method achieves optimal performance for incomplete multi-dimensional emotion recognition with modest computational overhead.

\section{Conclusions and future work}


To address feature selection in EEG-based emotion recognition with partially missing dimensional labels, ADSEL integrates adaptive dual self-expression and global redundancy learning within a least squares regression framework, enabling joint exploitation of sample and dimension information in the label space, thereby effectively leveraging both intra-sample and inter-dimension dependencies for robust label recovery. Experimental results show that ADSEL outperforms 14 popular methods, consistently achieving superior performance across all six evaluation metrics under various missing ratios. However, its label recovery performance is constrained under high missing rates and limited sample sizes, where the dual self-expression mechanism struggles to capture complex pairwise correlations among samples, dimensions, and their interactions, affecting recovery accuracy and generalization. Future work will focus on improving robustness to high missing rates and enhancing generalization in small-sample scenarios.

\bibliographystyle{IEEEtran}
{\footnotesize  
	\bibliography{IEEEabrv,myrefs}}


\end{document}